\documentclass[12pt]{article}
\usepackage{amsfonts,amsmath}
\usepackage{amssymb}

\usepackage[hypertex] {hyperref}

\newcommand{\dr}{{{\rm d}}}

\makeatletter \@addtoreset{equation}{section} \makeatother
\newcommand{\hhmt}{h}
\newcommand{\gt}{\tau}
 \newcommand{\eq}{\eqref}
\newcommand{\ie}{{\it i.e.,} }

{\vspace{3mm} }

\def\al{\alpha}

\def\*{\star}

\def\E2{\mathbf{E}}

\newcommand{\rhs}{{\it r.h.s.} }
\newcommand{\be}{\begin{equation}}
\newcommand{\ee}{\end{equation}}
\newcommand{\bee}{\begin{eqnarray}}
\newcommand{\beee}{\begin{array}}
\newcommand{\eee}{\end{eqnarray}}
\newcommand{\eeee}{\end{array}}
\newcommand{\ga}{\alpha}
\newcommand{\gb}{\beta}
\newcommand{\gga}{\gamma}

\newcommand{\ls}{\!\!\!\!\!\!}

\newcommand{\gd}{\delta}

\newcommand{\gep}{\epsilon}

\newcommand{\go}{\omega}

\newcommand{\dal}{\dot \alpha}
\newcommand{\dgb}{\dot \beta}

\newcommand{\q}{\,,\qquad}
\newcommand{\nn}{\nonumber}

\newcommand{\p}{\partial}

\newcommand{\ff}{\frac}
\newcommand{\hmt}{\vartriangle}

\begin{document}

\begin{flushright}
FIAN/TD/10-2018\\
\end{flushright}

\vspace{0.5cm}
\begin{center}
{\large\bf Homotopy Properties and Lower-Order Vertices in\\ Higher-Spin
Equations}

\vspace{1 cm}

\textbf{V.E.~Didenko${}^1$, O.A.~Gelfond${}^{1,2}$, A.V.~Korybut${}^1$ and  M.A.~Vasiliev${}^1$}\\

\vspace{1 cm}

\textbf{}\textbf{}\\
 \vspace{0.5cm}
 \textit{${}^1$ I.E. Tamm Department of Theoretical Physics,
Lebedev Physical Institute,}\\
 \textit{ Leninsky prospect 53, 119991, Moscow, Russia }\\

\vspace{0.7 cm}\textit{
${}^2$ Federal State Institution "Scientific Research Institute for System Analysis
of the Russian Academy of Science",}\\
\textit{Nakhimovsky prospect 36-1, 117218, Moscow, Russia}

\par\end{center}

\begin{center}
\vspace{0.6cm}
% didenko@lpi.ru, vasiliev@lpi.ru \\

\par\end{center}

\vspace{0.4cm}

\begin{abstract}
\noindent New homotopy approach to the analysis of nonlinear
higher-spin equations  is developed. It is shown to directly
reproduce the previously obtained local vertices. Simplest cubic
(quartic in Lagrangian nomenclature) higher-spin interaction
vertices in four dimensional theory are examined from locality
perspective by the new approach and shown to be local. The results
are obtained in a background independent fashion.
\end{abstract}
\newpage
\tableofcontents

\newpage
\section{Introduction}

The full nonlinear theory of interacting higher-spin (HS) gauge
fields  is currently known at the level of equations of motion
\cite{more}. At the action level only some lower-order results are
available (see {\it e.g.,}
\cite{Bengtsson:1983pd}-\cite{Francia:2016weg}). At cubic level,
however, the coupling constants are not  fixed by the Noether
procedure pretty much as the Yang-Mills coupling constants are
only get fixed by  Jacoby identities at the quartic order.
Nevertheless in  \cite{Fradkin:1987ks} the coupling constants were
fixed for the $4d$ vertices with spins $s_1,s_2,s_3$ obeying the
triangle inequality while in \cite{Metsaev:1993}  these results
were extended to any spins. Extension to any space-time dimension
was obtained in \cite{Sleight:2016dba} via
holographic computation. While  lacking conventional
action principle\footnote{The topological like AKSZ action that
reproduces full nonlinear HS equations in four dimensions was
constructed in \cite{BS}. To the moment it is not entirely clear
if it is suitable for standard QFT calculations.} prevents one
from treating the HS theory at the full fledged quantum level, the
Klebanov-Polyakov HS $AdS/CFT$ conjecture \cite{KP} (see also
\cite{Sezgin:2002rt}-\cite{SS}) points out to its quantum
consistency. The first nontrivial evidence of HS holographic
duality at tree level for certain three-point correlation
functions was presented by Giombi and Yin in \cite{GY1,GY2}. At
quantum level a remarkable cancellation of one loop determinant
divergencies was found in \cite{Giombi:2013fka,Beccaria:2014xda}.
It followed that the spectrum of HS fields itself essentially fine
tunes this cancellation.

A web of HS dualities relates in particular the simplest  free
$O(N)$ model   to a highly nontrivial $4d$ bulk HS theory. For
this particular case all boundary correlation functions in a
singlet sector are explicitly known \cite{GY1, Didenko:2012tv,
Gelfond:2013xt} and the HS bulk action can be reconstructed  order
by order using the boundary data. This program has been initiated
in \cite{Petkou:2003zz} and then further put forward in
\cite{Bekaert:2015tva} and \cite{Sleight:2016dba}. The resulting
local HS cubic action was shown to be consistent with the HS
algebra structure constants \cite{Sleight:2016xqq}. How this
method should be applied at higher orders is not quite clear since
already at quartic order the HS action is expected to be non-local
\cite{Bekaert:2015tva}, \cite{Sleight:2017pcz} while a part of
holographic reconstruction routine is a systematic discarding of
boundary terms which operation requires precise definition of the
non-local class of functions. Roughly speaking, the bulk terms can
be represented as boundary ones and vice versa whenever one is
free to use non-local operators of a kind $\ff{1}{\p}$. This makes
the problem of locality in HS theory of great importance.

One of the approaches on the way to understand (non-)locality in
HS theory is based on the analysis of holographic Mellin
amplitudes along the lines of \cite{Penedones:2010ue},
\cite{Bekaert:2015tva}, \cite{Rastelli:2017udc}. Despite some
difficulties in defining Mellin amplitudes for the free theory
correlators an encouraging result was obtained recently in
\cite{Ponomarev:2017qab} that seemingly singles out a specific
form of divergencies in quartic scalar interactions.

In this paper we  reconsider the problem within the bulk HS gauge
theory with no reference to holography. Specifically, by examining $4d$
HS equations perturbatively we find
restrictions imposed by locality on some simplest HS vertices.
Namely, we reconstruct interaction vertices which look
schematically
\begin{align}
&\dr_x\go=-\go*\go+\Upsilon(\go,\go,
C)+\Upsilon(\go,\go,C,C)+\dots\,,\label{ver1}\\
&\dr_x C=-[\go,C]_{*}+\Upsilon(\go,C,C)+\dots\label{ver2}
\end{align}
from HS equations of \cite{more}, where $\go(Y;K|x)$ and
$C(Y;K|x)$ are HS fields that apart from their dependence on
space-time coordinates $x$ depend also on auxiliary  spinor
variables $Y$ and Klein operators $K$. In doing so one faces
cohomological freedom of representatives in the spinor space
which may affect the form of $\Upsilon$-vertices and their
(non-)local behavior. Ideally, one would like to have  a
perturbative expansion procedure allowing to discard this
cohomological freedom yet having the resulting $\Upsilon$-vertices
within the proper locality class. Constructing such perturbation
theory is not trivial. Particularly, one of most natural choices
based on the {\it conventional homotopy} (see {\it e.g.,}
\cite{Vasiliev:Rev}) leads to non-local obstructions starting from
the second order \cite{GY1} (see also \cite{Boulanger:2015ova}).
At this order it was
shown in \cite{Vasiliev:2016xui,Gelfond:2017wrh} that up to local
freedom there exists a unique field redefinition that respects the
holomorphic factorization of the equations and results in local
cubic vertices on $AdS$ background. That these vertices do agree
with HS $AdS/CFT$ expectation was then  confirmed in
\cite{Sezgin:2017jgm}-\cite{Misuna:2017bjb}.

To extend the results of  \cite{Vasiliev:2016xui,Gelfond:2017wrh}
to higher perturbation orders it is necessary to elaborate a
systematic perturbative approach based on the homotopy techniques
different from the conventional one. In \cite{Gelfond:2018vmi} it was suggested
that the proper approach is based on the certain {\it shifted
homotopy} allowing to decrease the level of nonlocality as a
consequence of Pfaffian Locality Theorem (PLT) proven in that reference.
In this paper we further elaborate properties of shifted homotopy
operators applying them to the computation of perturbative
corrections and, in particular,  reproducing this way the previously obtained results
of \cite{Vasiliev:2016xui}.

The shifted homotopy technique is modified in two respects. First,
the shifted homotopy operators involve the shifts of arguments of
the dynamical HS fields. The number of such free parameters grows
with the order of perturbative expansion. The freedom in the field
variable choice due to these parameters is very limited confined
by a finite amount of free such parameters at a given perturbation
order. Compared to the generic field redefinitions that have
functional ambiguity the freedom in homotopy parameters represents
a very specific finite-dimensional subset. Second, we no longer
use the $AdS$ space as vacuum solution. The new technique allows
us to make the computation for general HS potentials $\go$ while
reconstructing the $C$--dependence order by order. From this point
of view vertex $\Upsilon(\go,\go,C)$ is the simplest cubic vertex
that shows up. We  confine ourselves to $\Upsilon(\go,\go, C)$ and
$\Upsilon(\go, C,C)$ and show how the free homotopy parameters
allow us to arrive at their local form. Let us stress that the
proposed approach is free from any kind of divergencies and
regularizations and
 needs no field redefinitions: the freedom in the choice of field variables is encoded in a few free
homotopy parameters.

We would also like to point out that the locality notion
considered in our paper might differ from the conventional
space-time locality. What we take as (non-)local is attributed to
spinor space rather than space-time, \ie to derivatives in
auxiliary spinor  rather than space-time variables. The two are
related via the unfolding routine that generally assumes
non-linear and infinite derivative one-to-one map. Therefore we
are dealing with spin locality rather than the space-time one (see
also discussion in \cite{Gelfond:2018vmi}). At the level
considered in our paper both notions are equivalent.   Our main
findings are the following.
\begin{itemize}

\item One of the main results of the paper is the development of the
useful properties of shifted homotopies of \cite{Gelfond:2018vmi}.
Remarkable feature of these homotopies is that they manifest a
striking interplay with HS star product relying on the
specific form of the latter.
No relations of this kind exist in the
conventional case being simply outside the box.

\item{{ While $\Upsilon(\go,\go, C)$ vertex is manifestly local
in any perturbative scheme we show that the homotopy parameters
can be chosen in such a way that it exhibits spin ultra-local
form. Namely, it reveals no dependence on (anti)holomorphic
auxiliary spinor variables in the $C$ field at all. This property
is a natural generalization of the central on-mass-shell theorem
\cite{Vasiliev:1988sa,Vasiliev:Rev} obtained for $AdS$ background.
The conventional homotopy belongs to this class. }} \item Using
the result of \cite{Gelfond:2018vmi} which says that the
conventional homotopy is not respected by PLT in the zero-form sector of
HS equations we extract vertex $\Upsilon(\go,C,C)$ using shifted
homotopies. We show that those (and only those) homotopy
parameters that are prescribed by the PLT of
\cite{Gelfond:2018vmi} generate local $\Upsilon(\go,C,C)$
generalizing that of \cite{Vasiliev:2016xui} obtained for $AdS$
background.

\item Finally, we show that though the admissible homotopies
result in equivalent local HS vertices up to some local field
redefinitions there is a one-parameter family of generalized
homotopies that reproduces identically equivalent
vertices $\Upsilon(\go,C,C)$ and  $\Upsilon(\go,\go,C)$.

\end{itemize}

{ Higher order vertices will be considered elsewhere. It will
be interesting to see if the vertex $\Upsilon(\go,\go,C,C)$ shares
spin ultra-local properties in the (anti)holomorphic sector or not. The result of
\cite{Gelfond:2017wrh} showing that (anti)holomorphic sector
vanishes on $AdS$ background points out that possibility.}

The paper is organized as follows. In section \ref{HSeq} we sketch
the locality problem and recall the HS equations as well as their
perturbative treatment. In section \ref{hmtp} we construct the
generalized homotopies and explore their properties. In section
\ref{pert} we further develop the perturbation theory based on the
shifted homotopies allowing us to compute HS vertices. Then in
section \ref{Gelfond} we briefly review PLT that constrains HS
vertices to their local form. In section \ref{vert} HS vertex
$\Upsilon(\go, C,C)$ is explicitly calculated. In section
\ref{locfr} we discuss homotopy shifts that produce local effects
at given order and in \ref{yshift} it is demonstrated that such
shifts may result in local field redefinitions. We conclude in
section \ref{con}. In Appendix it is demonstrated how homotopy
parameters that violate PLT bound result in non-local
$\Upsilon(\go, C,C)$  vertex.

\section{Higher-spin equations and locality}\label{HSeq}

Let us briefly recall the structure of HS equations. For more
detail we refer the reader to \cite{Vasiliev:Rev}. Within the
frame-like approach, HS dynamics is governed by one-form
$\go(y,\bar y;K|x)$ and zero-form $C(y, \bar y;K|x)$
 generated by all possible polynomials of $sp(4)$
spinors $Y_{A}=(y_{\al}, \bar{y}_{\dal})$, $\al,\dal=1,2$. HS
algebra is conveniently generated with the use of star product
\be
[y_{\al}, y_{\gb}]_*=2i\gep_{\al\gb}\,,\qquad [\bar{y}_{\dal},
\bar{y}_{\dgb}]_*=2i\gep_{\dal\dgb}\,,\qquad [y_{\al},
\bar{y}_{\dgb}]_*=0\,
\ee
defined as
\be\label{ystar}
f(y)*g(y)=f(y)e^{i\gep^{\al\gb}\overleftarrow{\p_{\al}}\overrightarrow{\p_{\gb}}}g(y)\,,
\ee
where $\gep_{\al\gb}$ and $\gep_{\dal\dgb}$ are two invariant
$sp(2)$ forms. Fields $\go$ and $C$ belong to different
representations of HS algebra distinguished by the extra
dependence on  the  Klein operators ${K}=(k,\bar k)$. Particularly, a spin $s$-field
is encoded by fields $\go_s$ and $C_s$ where $\go_s$ spans a
finite dimensional module of HS algebra as opposed to $C_s$ being
an infinite dimensional one. In other words, a given spin
$s$-field is stored in  polynomial $\go$ and unbounded $C$.
Components of $C(y,\bar y;K|x)$ are designed to contain HS Weyl
tensors along with all their on-shell nontrivial space-time
derivatives.

The precise field dependence on ${K}$ will be specified later on
and for now a schematic form of dynamical equations is given by
\eqref{ver1}, \eqref{ver2}. Their {\it r.h.s.\,}s say that HS
interactions are driven by HS algebra (first terms in \eqref{ver1}
and \eqref{ver2}) and its deformation which leads to HS vertices
$\Upsilon(\go, C\dots C)$ for the zero-form sector and
$\Upsilon(\go,\go, C\dots C)$  for one-forms. The form of these
vertices  can in principle be determined  from integrability
requirement $\dr_{x}^2=0$ up to field redefinition. In practice
however this kind of analysis gets increasingly complicated as the
order of $C$ grows \cite{Vasiliev:1988sa,Vasiliev:1989yr}.

Whatever these vertices are they stem from HS symmetry represented
by the star product. This is a source of possible
non-localities. Indeed, star product \eqref{ystar} is a non-local
operation in that it mixes any number of derivatives of two functions.
Still, vertices containing star product can be local.
As an example take the vertex $\go*\go$. Recall, that for a
given spin $s$ the corresponding $\go_s$ is a (degree $2(s-1)$ \cite{Vasiliev:Rev})
 polynomial in
$Y$'s. Product $\go_{s_1}*\go_{s_2}$ is also a polynomial and
therefore is local. Another cubic vertex entering \eqref{ver2} is
$[\go,C]_*$ which is local by similar argument. Despite
$C_s(y,\bar y|x)$ is not a polynomial, product $C_{s_1}*\go_{s_2}$
contain only finite amount of derivatives as follows from
\eqref{ystar}.
  Analogous reasoning brings one to a conclusion that
the simplest cubic vertex $\Upsilon(\go,\go, C)$ is local as well.
These exhaust the list of manifestly local vertices. Those
containing more than one $C$ are potentially non-local raising a
question of admissible class of functions that HS vertices should
belong to. At this stage two different potential situations are
not excluded. Either all $\Upsilon$-vertices are spin local for a given
set of spins entering the vertex or some include infinite number of
derivatives. To understand the structure of HS vertices we need to
proceed to generating HS equations that eventually lead to
\eqref{ver1}, \eqref{ver2}.

\subsection{Generating equations}

Dynamical HS equations \eqref{ver1}, \eqref{ver2} are reproduced
order by order from the following HS generating system \cite{more}
\begin{align}
&\dr_x W+W*W=0\,,\label{HS1}\\
&\dr_x S+W*S+S*W=0\,,\label{HS2}\\
&\dr_x B+[W,B]_*=0\,,\label{HS3}\\
&S*S=i(\theta^{A} \theta_{A}+\eta B*\gga
+\bar\eta B*\bar\gga)\,,\label{HS4}\\
&[S,B]_*=0\,.\label{HS5}
\end{align}
Here $W(Z,Y;{K}|x)$ is a one-form that eventually encodes
HS one-form $\go(Y;{K}|x)$ in \eqref{ver1}, \eqref{ver2}.
Apart from its dependence on generating variables $Y_{A}$, $W$
also depends on new auxiliary variables $Z_{A}=(z_{\al},
\bar{z}_{\dal})$. Together with $Y$'s these enhance spinor space
along with its star product
\be\label{star}
(f*g)(Z, Y)=\ff{1}{(2\pi)^4}\int dU dV f(Z+U, Y+U)g(Z-V,
Y+V)e^{iU_A V^A}\,.
\ee
For $Z$-independent functions \eqref{star} reduces  to
\eqref{ystar}. The following commutation relations can be easily
derived using \eqref{star}
\be
[Y_{A},Y_{B}]_*=-[Z_A,Z_B]_*=2i\gep_{AB}\,,\qquad [Y_A,Z_B]_*=0\,.
\ee
The fields $W$, $S$ and $B$ depend on a pair of Klein operators ${K}=(k,\bar k)$ that obey
\be\label{hcom}
\{k,y_{\al}\}=\{k,z_{\al}\}=0\,,\qquad [k,\bar y_{\dal}]=[k,\bar
z_{\dal}]=0\,,\qquad k^2=1\,.
\ee
Similarly for $\bar k$. In other words ($\bar k$)$k$ anticommutes
with (anti)holomorphic variables. Property $k^2=\bar k^2=1$
implies that field dependence on Klein operators is at most
bilinear, e.g., $W(Z,Y; K|x)=\sum_{m,n=0,1}W^{m,n}(Z,Y|x)k^m\bar
k^{n}$. Such dependence splits HS spectrum into propagating sector
and topological one \cite{Vasiliev:Rev}. The physical sector
is singled out by
\be\label{Wkl}
W(Z,Y; K|x)=W(Z,Y;-K|x)\,.
\ee
$B(Z,Y; K|x)$ is a zero-form  responsible for dynamics of $C$--
field \eqref{ver1}, \eqref{ver2}. $B$ depends on Klein operators
$k$ and $\bar k$, such that its propagating part enjoys
\be
B(Z,Y; K|x)=-B(Z,Y; -K|x)\,.
\ee

Another master field entering system \eqref{HS1}-\eqref{HS5} is
 $S(Z,Y; K|x)$. This field is purely auxiliary being
expressed via $C$ on-shell. As opposed to $W$, $S$  is a one-form
in additional direction spanned by anti-commuting differentials
$\theta_{A}=(\theta_{\al}, \bar\theta_{\dal})$ being a space-time
zero-form
\be
S=\theta^{\al}S_{\al}+\bar\theta^{\dal}\bar{S}_{\dal}\,,\qquad
\{\theta_{A},\theta_{B}\}=\{\theta_{A},\dr_x\}=0\,.
\ee
The
commutation rules for Klein operators are in accord with
prescription \eqref{hcom}
\be
\{\theta_{\al},k\}=\{\bar\theta_{\dal},\bar k\}=0\,,\qquad
[\theta_{\al},\bar k]=[\bar\theta_{\dal},k]=0\,.
\ee
Lastly, in the propagating sector of
the theory, the dependence of $S$ on Klein operators is similar to
\eqref{Wkl}
\be
S(Z,Y; K|x)=S(Z,Y; -K|x)\,.
\ee

Equation \eqref{HS4} contains the  central elements
\be\label{klein}
\gga=e^{iz_{\al}y^{\al}}k\theta^{\al}\theta_{\al}\,,\qquad
\bar\gga=e^{i\bar{z}_{\dal}\bar{y}^{\dal}}\bar
k\bar\theta^{\dal}\bar\theta_{\dal}
\ee
that commute with any element
$f(Z,Y;K;\theta; dx)$. This can be checked by noting using
\eqref{star} that
\be
e^{iz_{\al}y^{\al}}*f(z,\bar z, y,\bar y)=f(-z,\bar z, -y,\bar
y)*e^{iz_{\al}y^{\al}}\,,
\ee
similarly for $e^{i\bar{z}_{\dal}\bar{y}^{\dal}}$ and that
$\theta^3=\bar\theta^{3}=0$ due to two-component indices. Finally,
$\eta$ and $\bar\eta$ are the only free phase parameters
$(\eta\bar\eta=1)$ of equations \eqref{HS1}-\eqref{HS5}. Generally
they break parity of HS interaction unless $\eta=1$ or $\eta=i$ in
which cases the theory is called $A$-- and $B$--models
correspondingly \cite{SS}.

\subsection{Perturbative expansion: homotopy trick and field redefinitions}
To understand the way system \eqref{HS1}-\eqref{HS5} reproduces
\eqref{ver1}, \eqref{ver2} one  starts with perturbative
expansion. The proper vacuum for HS theory is  given by
\begin{align}
&B_0=0\,,\label{B0}\\
&S_0=\theta^A Z_{A}\,,\label{S0}\\
&W_0=\Omega=\ff
i4(\go_{\al\gb}y^{\al}y^{\gb}+\bar\go_{\dal\dgb}\bar y^{\dal}\bar
y^{\dgb}+2e_{\al\dal}y^{\al}\bar{y}{}^{\dal})\,.\label{W0}
\end{align}
 It is easy to check that
\eqref{HS1}-\eqref{HS5} is fulfilled provided $\go_{\al\gb}$,
$\bar\go_{\dal\dgb}$ and $e_{\al\dal}$ are $AdS_4$ connection
one-forms satisfying Cartan structure equations. Nonzero vacuum
value for $S$-field creates a pattern for finding $Z$-dependence
of fields. For example, at first order one has
\be
[S_0, B_1]_*+[S_1,B_0]_*=0\,.
\ee
Using \eqref{star} we have
\be
[S_0,f(Z,Y; K)]_*=-2i\theta^{A}\ff{\p}{\p Z^A}f=-2i\dr_Z f
\ee
and therefore $B_1$ is $Z$--independent
\be
B_1=C(Y; K)\,.
\ee
This is  the zero-form $C$ that appears in
\eqref{ver1}, \eqref{ver2}. Let us stress that it can be identified with the $\dr_Z$
cohomological part of $B$.

$S_1$ should be expressed in terms of
$C$. To find it one takes \eqref{HS4} from which we have
\be\label{S1}
-2i\dr_Z S_1=i\eta C*\gga+i\bar\eta C*\bar\gga\,.
\ee
This is a typical partial differential equation that determines HS
field $Z$-dependence. Terms on the {\it r.h.s.} are originated from
already given lower order contributions. Freedom in solutions to
such equations corresponds to the gauge and field redefinition
freedom.

A natural way of solving equation
\be\label{steq}
\mathrm{d}_{Z} f(Z;Y;\theta)=J(Z;Y;\theta)
\ee
with $\mathrm{d}_{Z} J=0$ is  by the homotopy trick. The simplest
choice of the homotopy operator for the exterior differential
$\mathrm{d}_{Z}$ is
\begin{equation}
\label{pQ}
\partial=(Z^{A} + Q^A)\frac{\partial}{\partial\theta^{A}}\,
\end{equation}
provided that
\be
\label{zind}
\frac{\p Q^B}{\p Z^A}=0\,.
\ee
Clearly, it obeys $\p^2=0$ (let us note that $Q^A$ can be an
operator). One has
\be
N:=\dr_Z \p +\p \dr_Z =
  \theta^{A}\frac{\partial}{\partial\theta^{A}}+(Z^{A}+Q^A)\frac{\partial}{\partial Z^{A}}\,.
\ee
Introducing the {\it almost inverse} operator
\be
   N^{*}g\left(Z;Y;\theta\right):=
 \intop_{0}^{1}dt\dfrac{1}{t}g\left(tZ -(1-t)Q;Y;t\theta\right)\label{Nhom}\q g(-Q;Y;0)=0
\ee
and the resolution operator
\be
\Delta_Q :=  \p N^{*} \q \Delta_Q g(Z;Y;\theta) =
(Z^{A}+Q^A)\dfrac{\partial}{\partial\theta^{A}}
\intop_{0}^{1}dt\dfrac{1}{t}g\left(tZ
-(1-t)Q;Y;t\theta\right)\label{homs}
\ee
gives the  resolution of identity
 \be\label{unit}
\{\dr_Z,\hmt_Q\}=1-h_Q\,,
\ee
where $h_Q$ is the following projector to the cohomology space
\be
\label{hf}
h_{Q}f(Z,\theta)=f(-Q,0)\,.
\ee

Resolution of identity allows one to write a particular solution to (\ref{steq}) in the
form
\be
\label{Q}
f_Q(J) = \hmt_Q J\,
\ee
provided that $h_Q J=0$ which condition is always true since
$J(Z;Y;\theta)$ in \eq{steq} is at least linear in $\theta^A$. General solution to (\ref{steq}) therefore
has the form
\be
\label{Qgen}
f(Z;Y;\theta) = f_Q(J)(Z;Y;\theta) +h(Y)+ \dr_Z \epsilon(Z;Y;\theta)\,,
\ee
where $h(Y)$ is in $\dr_Z$-cohomology while the last term is $\dr_Z$-exact.
Clearly, solutions (\ref{Q}) with different homotopy parameters $Q_1$ and $Q_2$ can differ
by a solution to the homogeneous equation, \ie
\be
\label{Q12}
f_{Q_1}(J)-f_{Q_2}(J)= h_{1,2}(Y)+ \dr_Z \epsilon_{1,2}(Z;Y;\theta)
\ee
with some $h_{1,2}(Y)$ and $\epsilon_{1,2}(Z;Y;\theta)$. (Explicit expressions for
$h_{1,2}(Y)$ and $\epsilon_{1,2}(Z;Y;\theta)$ follow from Eq.~\eq{hmtd}.)
In practical computations,
the addition of the cohomological terms $h_{1,2}(Y)$, that depend on the dynamical fields,
implies a (nonlinear) field redefinition. This is because the dynamical
fields $C(Y;K)$ and $\omega (Y;K)$ belong to the $\dr_Z$--cohomology. Hence, the homotopy choice in
\eq{Q} affects the choice of variables.
The  $\dr_Z \epsilon_{1,2}(Z;Y;\theta)$ term affects the gauge choice.

The choice of $Q=0$ corresponds to the conventional homotopy operator used in \cite{more}.
As explained below homotopies with non-zero shifts $Q$
 play important role in higher orders.

Proceeding further with the conventional resolution we  solve \eqref{S1} and
proceed to \eqref{HS2} that determines $Z$-dependence of $W_1$
\be\label{W1}
-2i\dr_Z W_1+D_{\Omega}S_1=0\,,\qquad
W_1=\go(Y;{K}|x)+\ff{1}{2i}\hmt_0D_{\Omega}S_1\,,
\ee
where $D_{\Omega}S_1:=\dr_x S_1+[\Omega, S_1]_*$ with the $AdS_4$
flat connection (\ref{W0}). Here $\go(Y;{K}|x)$ is the HS one-form
potential that appears in \eqref{ver1}, \eqref{ver2}. To obtain
first-order form of \eqref{ver2} one plugs $B_1$ into \eqref{HS3}
to have
\be
\dr_x C=-[\Omega,C]_*\,.
\ee
First-order form of \eqref{ver1} results from the  substitution of
\eqref{W1} into \eqref{HS1}
\be
\dr_x\go=-\{\Omega,
\go\}+\ff{i}{4}D_{\Omega}\hmt_0D_{\Omega}\hmt_0(\eta
C*\gga+\bar{\eta}C*\bar\gga)\,.
\ee
This way one finds linear contribution to
$\Upsilon(\go,\go, C)$ at $\go=\Omega$ having the form \cite{Vasiliev:Rev}
\begin{align}
\Upsilon(\Omega,\Omega,
C)=\ff{i}{4}D_{\Omega}\hmt_0D_{\Omega}\hmt_0(\eta
C*\gga+\bar{\eta}C*\bar\gga)=\\
=\ff i4\Big(\eta \bar H^{\dal\dal}\p^{2}_{\dal}\bar C(0, \bar y;
k, \bar k|x)\bar k+\bar\eta H^{\al\al}\p_{\al}^2 C(y, 0; k, \bar
k|x)k\Big)\,.\label{oms}
\end{align}
%where
%\be
%H_{\al\gb}=
%\ee

Higher order interactions can be extracted analogously.
The resulting vertices in
\eqref{ver1}, \eqref{ver2} are manifestly HS gauge covariant. A
systematic analysis of perturbation series based on conventional
homotopy  was elaborated in \cite{Didenko:2015cwv} (see also \cite{Sezgin:2002ru}).

It turns out however that starting from
 $\Upsilon(\Omega, C,C)$ application of conventional homotopy
results in infinite derivative tail and is
not consistent with locality. It was first observed in \cite{GY1}
that boundary correlation functions resulting from such a vertex
diverge. Later it was shown \cite{Vasiliev:2016xui} that there is
a field redefinition that respects the holomorphic factorization properties of
the solution, making this vertex local and consistent with
holographic limit \cite{Sezgin:2017jgm,Didenko:2017lsn}.

This fact suggests that
by considering a wider class of homotopy operators for
perturbative series one may hope to construct spin local or minimally non-local
HS interactions. In what follows we elaborate some
details of this programme.

\section{Homotopies and star product}\label{hmtp}

\subsection{Shifted homotopies}

In this section we analyze properties of shifted resolution
operators $\hmt_Q$ (\ref{homs}) starting with those
insensitive to the range of indices  of the variables $Z^A,Y^A, \theta^A$.

\subsubsection{General relations}
Operators $\hmt_{P}$ and $\hmt_{Q}$ anticommute,
\be\label{def1.0}
 \hmt_{P}\hmt_{Q}    = -\hmt_{Q}\hmt_{P}\,.
\ee
Indeed, by virtue of \eq{homs},
 \bee\label{delPdelQ1}
\Delta_P \Delta_Q f\left(  Z  ;Y; \theta\right) =\int_0^1 \dr \gt
\int_0^1  {\dr t}\,t
(Z^{B}+P^B) (\gt( Z^{A}+P^A) - P^A+Q^A) \times
\\ \nn  f_{AB}\left(t\gt Z -t(1-\gt)P
-(1-t)Q;Y;\gt t\theta\right)\,,
\eee
where
\be \label{fAB}f_{AB}\left(  Z  ;Y; \theta\right)
 :=  \dfrac{\partial^2}{\partial\theta^{A}\partial\theta^{B}}
f\left(  Z  ;Y; \theta\right)\,.
\ee
Using that
\be \label{symm}
\left(Z^B+P^B\right)\left(\tau\left(Z^A+P^A\right)-P^A+Q^A\right) f_{AB}=
\left(Z^B+P^B\right)\left(Z^A+Q^A\right) f_{AB}
\ee
since
 $f_{AB}=-f_{BA}$ and  changing the integration variables
\be\nn
\gt_3=\gt t\q \gt_2=1- t\q \gt_1=1-\gt_2-\gt_3
\ee
 \eqref{delPdelQ1} can be rewritten in the form

\be\label{delPdelQ}
\ls\hmt_P \hmt_Q f\left(  Z  ;Y; \theta\right) = \int_{[0,1]^3} {\dr^3
\tau\,} \gd(1\!-\!\tau_1\!-\!\tau_2\!-\!\tau_3)
(Z^{B}+P^B) (Z^A+Q^A)
%\times
%\\ \nn
f_{AB}\left( \gt_3 Z -  \gt_1 P
-\gt_2 Q;Y;\gt_3\theta\right)\,
\ee
making \eqref{def1.0} obvious.

Formula \eqref{delPdelQ} admits a neat generalization for
successive resolution operators
\bee\label{delPdelQn}
\hmt_{Q_n}\ldots \hmt_{Q_1} f\left(  Z  ;Y; \theta\right) =
\int_{[0,1]^{n+1}} {\dr^{\,n+1} \tau\,}
\gd\Big(1\!-\sum_{j=1}^{n+1}\tau_j \Big)\prod_{j=1}^{n}
\left(Z^{A_j}+ Q_j^{A_j}\right) \times \qquad\qquad\\ \nn
  \times f_{A_{n}\ldots A_{ 1}}
 \left( \gt_{n+1} Z -  \sum_{j=1}^n \gt_j Q_j
 ;Y;\gt_{n+1}\theta\right)\,\q
\eee
where\be\nn f_{A_{n} \ldots A_{ 1}}=
\dfrac{\partial}{\partial\theta^{A_{n }}}\ldots
 \dfrac{\partial}{\partial\theta^{A_1
 }}
f\left(  Z  ;Y; \theta\right)\,.
\ee
Also note that due to \eq{hf}
\bee\label{hdelPdelQn}
\hhmt_P\hmt_{Q_n}\ldots \hmt_{Q_1} f\left(  Z  ;Y; \theta\right) =
\int_{[0,1]^{n+1}} {\dr^{\,n+1} \tau\,}
\gd\Big(1\!-\sum_{j=1}^{n+1}\tau_j \Big) \prod_{j=1}^{n} \left(-
P^{A_j}+ Q_j^{A_j}\right) \times \qquad\qquad\\ \nn \times
f_{A_{n}\ldots A_{ 1}}
 \left( - \gt_{n+1} P -  \sum_{j=1}^n \gt_j Q_j
 ;Y;0\right)\,,
\eee
meaning that $\hhmt_P\hmt_{Q_n}\ldots \hmt_{Q_1}$ is totally
antisymmetric in indices $P$, $Q_j$. In particular,
\be
 \hhmt_{P}\hmt_{Q}    = -\hhmt_{Q}\hmt_{P} \,,
\ee
which has a consequence
\be
\label{PP}
 \hhmt_{P}\hmt_{P}    = 0\,.
\ee
Other useful  relations  are
\be\label{def1.2}
h_Ph_Q=h_Q\,,\qquad \hmt_P h_Q=0\,
\ee
and the consequence of  resolution of identity \eq{unit}
\be\label{hmtd}
 \hmt_B  -  \hmt_A =[\dr_Z,\hmt_A  \hmt_B]  +  \hhmt_A \hmt_B \,.\ee

In general in solving \eqref{steq} one is not confined to a
particular shift $Q$ in resolution operator $\hmt_Q$. Properly
normalized linear combination also gives a solution to
\eqref{steq}. One can take integrals over shift
parameters
\be
\hmt(\rho):= \int dQ \rho (Q) \hmt_Q
\ee
with the normalization condition
\be
\int dQ \rho (Q)=1\,.
\ee

\subsubsection{Two-component relations}
Now we restrict indicies of $Z_A,Y_A,\theta^A$ variables to take only two values
\begin{equation}
\left(Z_A,Y_A,\theta^A\right) \longrightarrow \left(z_\alpha,y_\alpha,\theta^\alpha\right)\,.
\end{equation}
The respective shifts will be denoted by lower case Latin letters.
In this case formulas \eqref{delPdelQn}, \eqref{hdelPdelQn} give
\bee\label{def1.0dok}
\hmt_b\hmt_a f(z,y) \theta^\gb\theta_\gb = 2 \int_{[0,1]^3}   {d^3
\gt\,} \gd(1\!-\!\gt_1\!-\!\gt_2\!-\!\gt_3)  (z+b)_\gga
 (z+a)^\gga
f( \gt_1 z-\gt_3 b-\gt_2 a ,y)\q
\quad\eee
\be\label{def1.0dok0}
\hhmt_c\hmt_b\hmt_a f(z,y) \theta^\gb\theta_\gb = 2 \int_{[0,1]^3}
{d^3 \tau\,} \gd(1\!-\!\tau_1\!-\!\tau_2\!-\!\tau_3)(b-c)_\gga
 ( a - c)^\gga f( -\tau_1 c-\tau_3 b-\tau_2 a ,y)\,.
\ee
From \eq{def1.0dok0} it follows in particular that
\be\label{h0}
h_{(\mu+1)q_2-\mu q_1}\hmt_{q_2}\hmt_{q_1}=0\q\forall \mu\in \mathbb{C}\,.
\ee
Applying \eq{def1.0dok}, \eq{def1.0dok0} to $\gga$  one finds
\be\label{ddg}
\hmt_b\hmt_a \gamma = 2\int_{[0,1]^3}   {d^3 \tau\,}
\gd(1-\tau_1-\tau_2-\tau_3)(z+b)_\gga
 ( z+a)^\gga e^{i (\tau_1 z -\tau_2 a-\tau_3 b)_{\al}y^\ga}k\q
\ee
\be\label{AdefHhhgam}\hhmt_{c}
   \hmt_{b} \hmt_{a}\gga
   =  2
\int_{[0,1]^3}   {d^3 \tau\,}
\gd(1-\tau_1-\tau_2-\tau_3)(b-c)_\gga
 (a-c)^\gga
e^{-i (\tau_1 c+\tau_2 a+\tau_3 b )_\ga y^\ga}k\,.
\ee
 Note that in accordance with \eqref{def1.0} the prefactor in
\eqref{ddg} is antisymmetric in  $a,b$  while the exponential is  symmetric.
Analogously, the \rhs of \eq{AdefHhhgam} is totally antisymmetric in $a,b,c$.
Also, from (\ref{AdefHhhgam}) it follows that
\be\label{idy}
h_{a+\al y}\hmt_{b+\al y}\hmt_{c+\al
y}\gga=h_{a}\hmt_{b}\hmt_{c}\gga\q \forall \al\in \mathbb{C}\,.
\ee

Let us also note that any homotopy containing $y$-shift  leaves no
effect on the exponential when applied to $\gga$. Indeed, from
\eqref{klein} and \eqref{homs} it follows that
\be\label{dg}
\hmt_{q+\al y}\gga=2(z^{\gb}+q^{\gb}+\al
y^{\gb})\theta_{\gb}\int_{0}^{1}dt t
e^{i(tz_{\al}-(1-t)q_{\al})y^{\al}}k
\ee
as parameter $\al$ drops out from the exponential being present in
prefactor. This property implies that the $y$-shifted homotopies
 do not affect locality at the order they are first applied.
  Note however that local field redefinition at a given order may
affect the structure of higher-order non-local vertices.

Finally, we need an identity that makes vertices \eqref{ver1} and
\eqref{ver2} resulting from generating equations
\eqref{HS1}-\eqref{HS5} manifestly $z$--independent. That this
should be so is granted by consistency of \eqref{HS1}-\eqref{HS5}.
The precise mechanism responsible for $z$-cancellation is stored
in identity resolution \eqref{unit} which makes some combinations
of shifted homotopies manifestly $z$--independent. A particularly
useful relation of this type is
\be\label{gammaid1}
(\hmt_d-\hmt_c)(\hmt_a-\hmt_b)\gga=(h_d-h_c)\hmt_a\hmt_b\gga\,,
\ee
which is true for any homotopy parameters $a,b,c,d$.  It can be
proven  as follows. First, using (\ref{hmtd}) along with
$\dr_z\gga=0$, $h_a\gga=0$, $h_a\hmt_b\gga=0$ one finds
$(\hmt_a-\hmt_b)\gga= \dr_z \hmt_b\hmt_a\gga$. Then, moving $\dr_z$ through
$(\hmt_d-\hmt_c)$ with the aid of identity resolution \eqref{unit} and using that
$ \hmt_a\hmt_b\hmt_c\gga =0$ $\forall a,b,c$ one obtains \eq{gammaid1}.

Eq.~\eqref{gammaid1} says in
particular that its {\it l.h.s.} is $z$-independent. Let us note that
identity resolution \eqref{unit} involves partial integration
which makes direct check of  \eqref{gammaid1} quite non-trivial.
As a simple consequence of \eqref{gammaid1} at $d=a$ one finds
\be \label{simJakobi}
   ( \hmt_c  \hmt_b -\hmt_c \hmt_a +\hmt_b  \hmt_a )\gamma=
 \hhmt_c   \hmt_b  \hmt_a   \gamma\,. \ee
Also applying $h_d$ to the both sides of this relation and using
\eq{def1.2} one obtains
\be \label{tri}
  \hhmt_d  \hmt_c  \hmt_b \gamma- \hhmt_d \hmt_c \hmt_a\gamma +\hhmt_d \hmt_b  \hmt_a \gamma=
 \hhmt_c   \hmt_b  \hmt_a   \gamma\,. \ee

In fact,  relation \eq{tri} expresses  {\it triangle identity} of \cite{Vasiliev:1989xz}
that played crucial role in the early analysis of nonlinear corrections to HS equations in \cite{Vasiliev:1988sa,Vasiliev:1989yr,Vasiliev:1990en} performed
with the help of the {\it triangle function}
\be
\label{tride}
\Delta(a_1,a_2,a_3) := \int_{[0,1]^3}   {d^3 \tau\,}
\gd(1-\tau_1-\tau_2-\tau_3)(a_{1\ga} a_2^\ga + a_{2\ga} a_3^\ga -a_{1\ga} a_3^\ga)\,
\delta^2\big (\sum_{i=1}^3 \tau_i a_i\big)\,
\ee
obeying triangle identity
\be
\label{trid}
\Delta(a,b,c) +\Delta (c,d,a) = \Delta(a,b,d) +\Delta (b,c,d)\,.
\ee

Comparing \eq{tride} with \eq{AdefHhhgam} we observe that $\hhmt_{c}
   \hmt_{b} \hmt_{a}\gga$ is a Fourier transform of $\Delta(a,b,c)$
\be\label{trf}
(\hhmt_{c} \hmt_{b} \hmt_{a}\gga)(y)\,
   =  2 \int d^2 u \Delta(c+u,b+u,a+u) e^{{i u_\ga y^\ga}}\, k\,
\ee
from where it is obvious that \eq{tri} is a consequence of
\eq{trid}. Note also that identity \eqref{idy} follows immediately
from \eqref{trf} upon the integration variable change $u_{\gb}\to
u_{\gb}+\al y_{\gb}$.

\subsection{Star-exchange homotopy relations}

One of the technical challenges in HS perturbation theory based on
conventional homotopy  is a permanent interplay between such
seemingly unrelated operations as star product  and homotopy
integration. This interplay renders algebraic structures obscure
and hinders one from constructing functional space that respects
both operations. Remarkably  shifted homotopies obey relations
that link the two operations as we now show. These properties do
not rely on particular range of indices of variables $z,y, \theta$  and
we take them as being generic in this section. The same time we discard the
right sector of dotted spinors that can be treated analogously.

Shifted homotopy operators obey remarkable {\it star-exchange}
relations with $z$-independent star-product elements. Namely, let
us consider a homotopy action on a star product
$C(y;k)*\phi(z,y;k;\theta)$.
  Using \eqref{star} and
\eqref{homs} one can check that
\be\label{lpr}
\hmt_{q+\al y}
(C(y;k)*\phi(z,y;k;\theta))=C(y;k)*\hmt_{q+(1-\al)p+\al
y}\phi(z,y;k;\theta)\,,
\ee
where $q$ is a $y$--independent parameter and $\al$ is a number.
We have also introduced the notation
\be\label{p}
 p_{\al}C(y;k)\equiv
C(y;k)p_{\al}:=-i\ff{\p}{\p y^{\al}}(C_{1}(y)+C_{2}(y)k)\,,
\ee
where $C(y;k)=C_{1}(y)+C_{2}(y)k$. Note that $p$ acts on the argument of the
$z$--independent function only which in our case is $C$ no matter
if it appears on the left or on the right. The
$z$-independence of $C$ is crucial for \eqref{lpr} to take place.
Let us stress also, that though $C=C(y;k)$ depends on $k$, operator
$p_{\al}$ is defined to commute with $k$, $[p_{\al},k]=0$, and therefore
is insensible to such dependence. Its action on $C$ by our
definition can be either written from left or right \eqref{p}.

While it is not difficult to check \eqref{lpr} directly using
\eqref{star} and \eqref{homs}, in doing so one encounters certain
cancellations in star product computation that come along
simultaneously both in the exponential and pre-exponential which
may seem coincidental. Still, eq. \eqref{lpr} is not that
surprising after all as can be seen from the following
consideration. Suppose one is to solve the equation
\be\label{eq}
\dr_z f=C(y;k)*\phi(z,y;k;\theta)\,
\ee
with $\dr_Z$-closed $\phi(z,y;k;\theta)$.
One way to do it is by using \eqref{homs} which gives
$f=\hmt_{a}(C(y;k)*\phi(z,y;k;\theta))$ with some homotopy
parameter $a$. On the other hand,
since $C(y;k)$ is $z$--independent and hence commutes with $\dr_z$,
 one can solve \eqref{eq} in the
form $f=C(y;k)*\hmt_{b}\phi(z,y;k; \theta)$ with some other parameter $b$
thus suggesting that $a$ and $b$ should be related. The precise
relation is given by \eqref{lpr}. This reasoning makes it clear
why $C(y;k)$ should be $z$--independent to yield \eqref{lpr}.

Analogously, it can be shown that
\bee\label{rpr}
\hmt_{q+\al y}(\phi(z,y ;\theta)*k^\nu *C(y;k))&=&\hmt_{q+ (-1)^\nu(1+\al)p+\al
y}\big(\phi(z,y; \theta)*k^\nu\big)*C(y;k)\\ \nn&=&\hmt_{q+ (-1)^\nu(1+\al)p+\al
y}\big(\phi(z,y; \theta)\big)*k^\nu*C(y;k)\,.
\eee
Note that the additional sign factor $(-1)^\nu$ is due to the definition \eq{p} of the shift $p$ as
derivative over the total argument of $C(y;k)$.

 There are two special points $\al=\pm 1$ related
to the normal ordering of star-product variables
$y\pm z$. Right product \eqref{rpr} remains invariant for
$\al=-1$, whereas left one \eqref{lpr} for $\al=1$.

Since any perturbative correction depends on star products of the
$Z$-independent fields $\go(Y;K)$ and $C(Y;K)$ using star-exchange
formulas any perturbative result can be reduced to the form where
shifted homotopy operators act only on the central two-form
elements $\gamma$ and $\bar\gamma$
 \eqref{klein}.
Since $\gga$ is central, one is able to calculate
$C*\hmt_q\gga\to\hmt_{q'}\gga*C$ using for example \eqref{lpr}. Taking
into account that $\gga$ is linear in $k$ (analogously for
$\bar\gga$) this results in the following rule
\be\label{drag}
\hmt_{\tilde q}\gga*C(y;k)=C(y;k)*\hmt_{\tilde q+2p}\gga\q
\tilde q = q+\al y\,.
\ee
 Formula \eqref{drag}
will be used steadily in our analysis of HS vertices.

Analogously it is straightforward to check the following useful
identities with $y$-independent $q$:
\bee
\hhmt_{q+\al y}
(C(y;k)*\phi(z,y;k;\theta))\!&=&\!C(y;k)*\hhmt_{q+(1-\al)p+\al
y}\phi(z,y;k;\theta)\,,\label{lprh}
\\
\hhmt_{q+\al y}(\phi(z,y ;\theta)*k^\nu *C(y;k))\!&=&\!\hhmt_{q+ (-1)^\nu(1+\al)p+\al
y}\big(\phi(z,y; \theta)*k^\nu\big)*C(y;k) \label{rprh}\,.
\eee

It should be  stressed that formulae presented in this section heavily rely on the specific form of
star product  (\ref{star}).
An important consequence of our analysis  is that the class of
homotopies with shift parameters $Q$  (\ref{pQ}) linear in derivatives $P_A$
of the $Z$-independent
fields $C(Y;K)$ and $\omega(Y;K)$ and/or $Y$-variables, that obviously obey the $Z$-independence
 condition (\ref{zind}), remains closed  under the
 star-product exchange formulas. It is this class of {\it linear shifted homotopies} that
 turns out to be most appropriate for the perturbative analysis of HS equations.

\section{Lower-order vertices}\label{pert}

The transition from one homotopy to another
not only affects the gauge choice in the $Z$-space ({\it i.e.,} $\dr_Z$-exact terms) but also the cohomology
representative as is most obvious from (\ref{Q12}). The latter freedom encodes different choices of
field variables. Hence, the proper choice of the homotopy operators can lead directly to the
local result with no need of  further field redefinitions. In the rest of this paper we demonstrate
 how this works in practice.

Using generalized resolutions \eqref{homs} we can
 step away from conventional homotopy approach and
reconsider perturbative analysis of \eqref{HS1}-\eqref{HS5}.
Recall that within the standard approach one starts with the $AdS$
background as vacuum solution \eqref{B0}-\eqref{W0}. Identities
\eqref{lpr} and \eqref{rpr} actually allow us to develop
perturbative expansion for any HS one-form $\go(Y;{K})$
reproducing directly vertices \eqref{ver1}, \eqref{ver2} as
$C$-expansion in spirit of \cite{Vasiliev:1988sa}.

\subsection{$\Upsilon(\go,\go,C)$ -- vertex}
\label{ooc}
%\subsubsection{Conventional homotopy}
At first order on $AdS$ we know that conventional homotopy works
fine reproducing canonical form \eqref{oms} that identifies HS
Weyl tensors with derivatives of HS potentials. We may now redo
this analysis for generic $\go$ thus reproducing the entire vertex
$\Upsilon(\go,\go,C)$. Let us start with \eqref{S1} from which we
find
\be\label{S1om}
S_1=-\ff{\eta}{2}\hmt_0(C*\gga)+c.c.=-\ff{\eta}{2}C*\hmt_{p}\gga+c.c.\,.
\ee

Here we made use of \eqref{lpr}. (Recall that index $p$ denotes
differentiation \eqref{p}.) From \eqref{HS2} we then have
\be\label{W1conv}
W_{1}=\ff{1}{2i}\hmt_{0}(\dr_x S_{1}+\go *S_1 +S_1 *\go)+c.c.\,.
\ee
Substituting here \eqref{S1om} and using
\be
\dr_x C+[\go,C]_*=0
\ee
and \eqref{rpr} one gets
\be\label{W1om}
W_1=-\ff{\eta}{4i}\left( C*\go*\hmt_{p+t}\hmt_{p+2t}\gga-
\go*C*\hmt_{p+t}\hmt_{p}\gga\right)+c.c.\,,
\ee
where $t_{\al}$ acts on $\go$
\be
 t_{\al}\go(Y;K):=-i\ff{\p}{\p
y^{\al}}\go(Y;K)\,.
\ee
To obtain
$\Upsilon(\go,\go,C)$ it remains to plug \eqref{W1om} into
\eqref{HS1} which after some elementary algebra making use of \eqref{simJakobi} and \eq{PP} gives
\be\label{go1}
\dr_x\go+\go*\go=\ff{\eta}{4i}(\go*\go*C*X_{\go\go
C}+C*\go*\go*X_{C\go\go}+\go*C*\go*X_{\go C\go})+c.c.\,,
\ee
where
\begin{align}
&X_{\go\go C}=h_{p+t_1+t_2}\hmt_{p}\hmt_{p+t_2
}\gga\,,\label{X1}\\
&X_{C\go\go}=h_{p+t_1+t_2}\hmt_{p+t_1+2t_2}\hmt_{p+2t_1+2t_2
}\gga\,,\label{X2}\\
&X_{\go C\go}=-h_{p+t_1+t_2}\hmt_{p+t_1+2t_2}\hmt_{p+t_2}\gga
-h_{p+t_1+2t_2}\hmt_{p+2t_2}\hmt_{p+t_2}\gga\,.\label{X3}
\end{align}
Carrying out star-product integration on the \rhs of
 \eqref{go1} and using \eqref{AdefHhhgam} we find
\be
\Upsilon(\go,\go,C)=\Upsilon_{\go\go
C}+\Upsilon_{C\go\go}+\Upsilon_{\go C\go}\,,
\ee
where
\begin{align}\label{gogoC}
&\Upsilon_{\go\go
C}=\ff{\eta}{2i}\int_{[0,1]^3}d^3\tau\gd(1-\tau_1-\tau_2-\tau_3)
e^{i(1-\tau_3)\p_{1}^{\al}\p_{2\al}}\\
&\p^{\al}\go((1-\tau_1)y,\bar
y)\,\overline{*}\,\p_{\al}\go(\tau_2y,\bar y)\,\overline{*}\,
C(-i\tau_1\p_1-i(1-\tau_2)\p_2,\bar y; K)k\,,\notag
\end{align}
\begin{align}\label{Cgogo}
&\Upsilon_{C\go\go
}=\ff{\eta}{2i}\int_{[0,1]^3}d^3\tau\gd(1-\tau_1-\tau_2-\tau_3)
e^{i(1-\tau_3)\p_{1}^{\al}\p_{2\al}}\\
&C(i\tau_1\p_2+i(1-\tau_2)\p_1,\bar y;
K)\,\overline{*}\,\p^{\al}\go(\tau_2y,\bar
y)\,\overline{*}\,\p_{\al}\go(-(1-\tau_1)y,\bar y)k \,,\notag
\end{align}
\begin{align}\label{goCgo}
&\Upsilon_{\go C\go
}=\ff{\eta}{2i}\int_{[0,1]^3}d^3\tau\gd(1-\tau_1-\tau_2-\tau_3)
e^{i(1-\tau_3)\p_{1}^{\al}\p_{2\al}}\\
&\p^{\al}\go(\tau_1y,\bar
y)\,\overline{*}\,C(i(1-\tau_2)\p_2-i(1-\tau_1)\p_1,\bar
y; K)\,\overline{*}\,\p_{\al}\go(-(1-\tau_2)y,\bar y)k+\notag\\
&+\ff{\eta}{2i}\int_{[0,1]^3}d^3\tau\gd(1-\tau_1-\tau_2-\tau_3)
e^{-i\tau_2\p_{1}^{\al}\p_{2\al}}\notag\\
&\p^{\al}\go((1-\tau_1)y,\bar
y)\,\overline{*}\,C(-i\tau_1\p_1+i\tau_3\p_2,\bar y;
K)\,\overline{*}\,\p_{\al}\go(-(1-\tau_3)y,\bar y)k\,.\notag
\end{align}
Here $\overline{*}$  denotes the star product with respect to the
leftover $\bar y$ variables and $\p_{1\al}$ and $\p_{2\al}$ are
differentiations of first and second $\go$'s correspondingly
counted from left to right. Remarkable property of the obtained
vertex is that all its structures contain only  derivatives of
$C(0,\bar y;K)$ having no dependence on $y$ in $C(y,\bar y;K)$ at
all. This fact is not accidental being a consequence of PLT as
explained in the next section. Such a form of deformation
effectively makes spin--$s$ contribution from $C$ finite component
for any given $\go$. On $AdS$ background, where $\go=\Omega$
\eq{W0}, the vertex is given by \eqref{oms} and contains $\p\p
C(0,\bar y;K)$. Eqs. \eqref{gogoC}-\eqref{goCgo} generalize this
property to cubic (quartic order in the Lagrangian nomenclature).
{A local vertex that contains at most a finite number of derivatives
of $C$'s at $y=0$ we call {\it ultra-local}. While the locality of
$\Upsilon(\go,\go,C)$ is granted in any perturbative scheme, the
fact that it appears in a ultra-local form can be crucial for
higher-order locality. }

\subsection{Classes of functions and Pfaffian Locality Theorem }\label{Gelfond}
As already mentioned,  at the level $O(C^2)$ conventional homotopy leads to a  non-local
result for, say, $\Upsilon(\Omega, C,C)$. The non-locality results
from the application of the conventional homotopy for the
reconstruction of the $z$-dependence of $B$ field from \eqref{HS5}
and upon substitution it to the physical sector equation
\eqref{HS3} yields $\Upsilon(\go,C,C)$ which appears non-local
even on $AdS$, \ie at $\go=\Omega$. The origin of this phenomenon is as
follows. Suppose one solves for $z$-dependence of $W$ and $S$
using conventional homotopy  all the way. This results in that
star products $S*S$, $W*W$, $W*S$, $S*W$ belong to one and the
same class of functions found in \cite{Gelfond:2018vmi} that we soon specify. Now, if one to solve
for $B$ using conventional homotopy then $W*B$, $B*W$, $B*S$,
$S*B$ belong to the same class of functions while $B*\gga$ does
not. That is a product with inner Klein operator
$e^{iz_{\al}y^{\al}}$ ruins the aforementioned class of functions.
A way out is to use different homotopy resolution \eqref{homs} for
$B$ field. This turns out to be possible and leads to consistent
classes of functions for both zero-forms and one-forms. The fact
that one-forms are reconstructed using the conventional homotopy
severely restricts homotopy choice for zero-form. More generally,
in HS systems containing higher-rank differential forms as in \cite{Vasiliev:2015mka}
homotopy operators in perturbation theory are expected to be $\mathbb{Z}_2$-graded with respect
to the rank of differential forms.
The condition on homotopy in the zero-form sector is governed by PLT
 of \cite{Gelfond:2018vmi}.

Here we sketch the Structure Lemma and PLT of \cite{Gelfond:2018vmi} which
prescribe classes of homotopies for one- and zero-form holomorphic sectors of
HS equations \eqref{HS1}-\eqref{HS5}. Remarkably, for these
classes PLT reduces the degree of non-locality making the
considered in this paper vertices manifestly local. Leaving the
reader with \cite{Gelfond:2018vmi} for details, we would like to
illustrate the logic using the example of conventional homotopy.

Confining ourselves to the holomorphic sector, it is convenient to work with Taylor expanded fields
$\go(y)=\go(0)e^{y^{\al}\p_{\al}}$ and
$C(y)=C(0)e^{y^{\al}\p_{\al}}$. Here and below in this section
differentiation $\p^{i}_{\al}$ with respect to $i^{th}$ argument
acts leftwise and commutes with $k$. Star-product
 \eqref{ystar} of two fields contains a factor
\be
e^{y^{\al}\p_{1\al}}*e^{y^{\al}\p_{2\al}}=e^{y^{\al}(\p_{1\al}+\p_{2\al})-i\p_{1}^{\al}\p_{2\al}}
\ee
with potentially nonlocal contribution
$e^{-i\p_{1}^{\al}\p_{2\al}}$. Apart from star products,
Eqs.~\eqref{HS1}-\eqref{HS5} contain inner Klein operator
$e^{iz_{\al}y^{\al}}$. Perturbative expansion includes also
homotopy operator  actions \eqref{homs} involving
both $y$'s and $\p$'s. All these building blocks together imply
the following form for, say, $O(C^n)$ perturbative contribution

\be
\label{exp}
\int\dots\int_{[0,1]^n} dt^nC(0)\dots C(0) e^{i(T
z_{\al}y^{\al}-i A^{j}\p^{\al}_{j}z_{\al}-iB^{j}\p^{\al}_{j}y_{\al}-\ff12P^{ij}\p^{\al}_{i}\p_{\al
j})}\,,
\ee
where $T$, $A^i$, $B^i$ and $P^{ij}$ depend on $n$ homotopy
integration parameters $t_1,\dots, t_n$. Star product of
such two contributions of $n$-th and $n'$-th orders gives
\begin{align}
&e^{iT
z_{\al}y^{\al}+A^{j}\p^{\al}_{j}z_{\al}+B^{j}\p^{\al}_{j}y_{\al}-\ff
i2P^{ij}\p^{\al}_{i}\p_{\al j}}* e^{iT'
z_{\al}y^{\al}+A^{i'}\p^{\al}_{i'}z^{\al}+B^{i'}\p^{\al}_{i'}y_{\al}-\ff
i2P^{i'j'}\p^{\al}_{i'}\p_{\al
j'}}= \\
&e^{i(T\circ T')
z_{\al}y^{\al}+A^{i''}\p^{\al}_{i''}z_{\al}+B^{i''}\p^{\al}_{i''}y_{\al}-\ff
i2P^{i''j''}\p^{\al}_{i''}\p_{\al j''}}\,,
\end{align}
where
\begin{align}
&T\circ T'=T+T'-2T T'\,,\label{p1}\\
&A^{i''}=(1-T')A^{i}+(1-T)A^{i'}+TB^{i'}-T'B^{i}\,,\label{p2}\\
&B^{i''}=(1-T')B^{i}+(1-T)B^{i'}+TA^{i'}-T'A^{i}\,,\label{p3}\\
&P^{i''j''}=P^{ij}+P^{i'j'}+(A^i+B^i)(A^{j'}-B^{j'})-(A^j-B^j)(A^{i'}+B^{i'})\label{p4}\,.
\end{align}

A striking property of star product \eqref{star} which follows
from \eqref{p1}-\eqref{p4} is as follows. Suppose one solves for
field $z$-dependence using conventional homotopy  all the way in
perturbation then, as shown in \cite{Gelfond:2018vmi},
coefficients  $T$, $A$, $B$ and $P$ satisfy\footnote{From \eq{B}
it follows that $B_i=0$ if $i$ takes only one value.  Hence, in
accordance with the analysis of section
 \ref{ooc}, from here it follows that the correction in the sector linear
in the zero-forms $C$ is $y$-independent being ultra-local provided that
one stays in the proper class.}
\begin{align}
&\sum(-)^j A^j=-T\,,\label{A}\\
&\sum(-)^j B^j=0\,,\label{B}\\
&\sum(-)^iP^{ij}=B^j\label{P}\,.
\end{align}
(Sign alternation $(-)^j$ appears
due to the presence of the outer Klein operator
$k$ in $\gamma$ \eq{klein}.) This property is also invariant under star product \eqref{star}. For instance, if $S$
field at some orders $S^i$ and $S^{i'}$ satisfy
\eqref{A}-\eqref{P} then so does $S^{i''}=S^{i}*S^{i'}$.

Looking at \eqref{HS4} we see that it has $S*S$ term which belongs
to  \eqref{A}-\eqref{P} as soon as one uses conventional homotopy.
Another contribution to \eqref{HS4} is $B*\gga$.  If $B$ is solved
by using the conventional homotopy then $B*\gga$ does not belong
to the class \eqref{A}-\eqref{P} simply because the star product
of $B$ with $\gga$ swaps $z$ and $y$. However, it turns out
that a proper shifted homotopy exists that brings $B*\gga$ to the
class \eqref{A}-\eqref{P}. Indeed, in
order $B*\gga$ to respect \eqref{A}-\eqref{P}
\be
e^{iT
z_{\al}y^{\al}+A^{i}\p^{\al}_{i}z_{\al}+B^{i}\p^{\al}_{i}y_{\al}+\ff
i2P^{ij}\p^{\al}_{i}\p_{\al
j}}*e^{iz_{\al}y^{\al}}=e^{i(1-T)z_{\al}y^{\al}-A^{i}\p^{\al}_{i}y_{\al}-B^{i}\p^{\al}_{i}z_{\al}+\ff
i2P^{ij}\p^{\al}_{i}\p_{\al j}}
\ee
the following conditions
\begin{align}
&\sum(-)^j A^j=0\,,\label{B-A}\\
&\sum(-)^j B^j=1-T\,,\label{B-B}\\
&\sum(-)^iP^{ij}=-A^j\label{B-P}
\end{align}
should be imposed. Eqs. \eqref{B-A}-\eqref{B-P} place strong
constraints on $B$-field. Indeed, as $B^{i''}$ originates from
homotopy acting on $S^{i}*B^{i'}$ where the two fields belong to
different classes \eqref{A}-\eqref{P} and \eqref{B-A}-\eqref{B-P}
the fact that $B^{i''}$ remains in \eqref{B-A}-\eqref{B-P} is
non-trivial and requires specific homotopy. Let us take most
general homotopy up to $y$-shifts (local effects) to see if it is
capable to reconcile the result with \eqref{A}-\eqref{P}
\be\label{hom}
z_{\al}\to z_{\al}-i\sum v^{i''}\p_{\al i''}\,,
\ee
where $v^{i''}$ are some numbers. Performing star product of
$S^{i}$ and $B^{i'}$ and applying homotopy \eqref{hom} to the
obtained result entails the following coefficients in $B''$
\begin{align}
&A^{i''}=T''\big((1-T')A^i+ (1-T)A^{i'}+ T
B^{i'}- T'B^i\big)\,,\\
&B^{i''}=(1-T')B^{i}+(1-T)B^{i'}+T A^{i'}-T'
A^{i}+(1-T'')(T\circ
T')v^{i''}\,,\\
&P^{i''j''}=P^{ij}+P^{i'j'}+(A^i+B^i)
(A^{j'}-B^{j'})-(A^{j}-B^{j})(A^{i'}+B^{i'})-\ff{1-T''}{T''}(A^{i''}
v^{j''}-A^{j''}v^{i''})\,,
\end{align}
where $T$, $T'$ and $T''$ are the homotopy integration variables
entering in $S$, $B'$ and $B''$ correspondingly. Assuming now that
dashed and double-dashed fields are from \eqref{B-A}-\eqref{B-P}
and non-dashed are from \eqref{A}-\eqref{P} we find following
\cite{Gelfond:2018vmi}
\be\label{homb}
\sum (-)^{j''}v^{j''}=1\,.
\ee
Thus the homotopy prescription used for one-forms ($S$ and $W$)
requires modified homotopy \eqref{hom}, \eqref{homb} for zero-form
$B$ in order to preserve functional classes \eqref{A}-\eqref{P}
for one-forms and \eqref{B-A}-\eqref{B-P} for zero-forms. A short
explanation to this phenomenon is the presence of inner Klein
operator in $\gga$ which comes with $B$-field only, affecting the
structure of the perturbative expansion. The effect it brings is
compensated by modified homotopy. That such a compensation
mechanism works at all at the condition \eqref{homb}  is highly non-trivial
and is related to a very special form of star product
\eqref{star}. Clearly the conventional homotopy $v^{i''}=0$
violates \eqref{homb}.

Another important fact shown in \cite{Gelfond:2018vmi} is that
Structure Lemma and, hence, PLT
 is respected by the $\dr_x$ differential which may have a
nontrivial effect for shifted homotopies. The fact that shifted homotopies
obeying conditions of PLT decrease the level of nonlocality
follows from the $Z$-dominance Lemma  proven in \cite{Gelfond:2018vmi}.

At second order $O(CC)$ the number of free parameters in
\eqref{hom} when solving for $B$ is two. Condition \eqref{homb}
reduces it further to a one-parameter family
\be\label{C2}
v_2-v_1=1\,.
\ee
The resulting contribution based on homotopy \eqref{C2} to the
cubic vertex $\Upsilon(\go, C,C)$ turns out to be local
generalizing the known cubic vertex $\Upsilon(\Omega, C,C)$  in $AdS_4$
\cite{Vasiliev:2016xui}. Moreover the one-parameter freedom
\eqref{C2} appears to be  spurious dropping off from the final
result as explained in section \ref{vert}. We now analyze
$O(C^2)$-type vertices in more detail.

\subsection{$\Upsilon(\go,C,C)$ -- vertex}\label{vert}
%\subsubsection{$Y$-independent shifts}
Let us start  by sketching how \eqref{homb} leads to correct local
vertex $\Upsilon(\Omega, C,C)$. To solve for $B$ to the second
order we take $S_1$ (\ref{S1om}) giving
\be
S_{1}=\theta^{\al}S_{1\al}+c.c.\,\qquad
S_{1\al}=\eta\int_{0}^{1}dt tz_{\al}C(-tz, \bar
y;K)e^{itz_{\al}y^{\al}}k
\ee
and from \eqref{HS5} have
\be\label{B2}
[S_0, B_2]+[S_1, B_1]=0\quad\Rightarrow\quad
2i\theta^{\al}\ff{\p}{\p z^{\al}} B_2=S_1*C-C*S_1\,,
\ee
which results in
\be
2i\ff{\p}{\p z^{\al}}B_2=A_{\al}-B_{\al}\,,
\ee
where
\begin{align}
&A_{\al}=\eta\int_{0}^{1}
t(z_{\al}-i\p_{2\al})e^{it(z_{\al}-i\p_{2\al})(y^{\al}-i\p_{2}^{\al})-t(z^{\al}-i\p_{2}^{\al})\p_{1\al}-y^{\al}\p_{2\al}}
C(0,\bar y; K)\,\overline{*}\,C(0,\bar y; K)k\,,\\
&B_{\al}=\eta\int_{0}^{1}
t(z_{\al}+i\p_{1\al})e^{it(z_{\al}-i\p_{1\al})(y^{\al}+i\p_{1}^{\al})+y^{\al}\p_{1\al}+t(i\p_{1}^{\al}-z^{\al})\p_{2\al}}
C(0,\bar y; K)\,\overline{*}\,C(0,\bar y; K)k\,.
\end{align}
We are ignoring $\bar\eta$ terms for brevity and $\bar*$ denotes
the star product with respect to the right variables $\bar
y^{\dot\alpha}$. Now we can solve for $B_2$ using shifted homotopy as follows
\be
2iB_2=\hmt_{q} (\theta^{\al}(A_{\al}-B_{\al}))\q \label{hommn}
q=v_1p_1+v_2p_2 \,,
\ee
 where $v_1$ and $v_2$ are some numbers.
Imposing homotopy condition \eqref{C2}, after some algebra
involving partial integration one can obtain
\begin{equation}
\label{der} B_2=B_{2\eta}^{loc} -\frac{\eta}{2}\int_0^1 dt
\,C\left(ty,\bar{y}; K\right)\,\overline{*}\,C\left(\left(t-1\right)y,\bar{y};
K\right)k\,,
\end{equation}
where
\begin{multline}\label{B2vas}
B_{2\eta}^{loc}=\ff{\eta}{2}\int_{[0,1]^3}d^3\tau\left(\gd'(X)-iz_{\al}y^{\al}\gd(X)\right)
e^{\tau_3z_{\al}(\p^{\al}_2+\p^{\al}_{1})+i\tau_3\p_{1\al}\p^{\al}_{2}+i\tau_3z_{\al}y^{\al}-y^{\al}(\tau_2\p_{2\al}-\tau_1\p_{1\al})}\times\\
\times C(0,\bar{y}; K)\,\overline{*}\,C(0,\bar{y}; K)k\,,
\end{multline}
\begin{align}
X=1-\tau_1-\tau_2-\tau_3\,,
\end{align}
which turns out  independent of the remaining parameter $v_1+v_2$. Here $B_{2\eta}^{loc}$ coincides
 with that found in \cite{Vasiliev:2017cae} which
is known to reproduce correct local $\Upsilon(\Omega,C,C)$ in $AdS$ background.
Let us stress that  our
approach is free of any field redefinitions thanks to the proper homotopy  choice leading directly
to the local result. The resulting $B_2$ differs from $B_{2\eta}^{loc}$ by the second term
on the \rhs of (\ref{der}) which, containing no contractions between first arguments
of the two factors of $C$, is local.

After this old fashion-style exercise we proceed to a more systematic analysis of the
 vertex $\Upsilon(\go,C,C)$. We want to obtain \eqref{ver2}
up to terms quadratic in $C$. To do that one solves \eqref{HS5} up
to this order. Namely, from \eqref{B2} and \eqref{S1om} we have
(ignoring $\bar\eta$--terms)
\be
-2i\dr_z
B_2=-\ff{\eta}{2}C*C*\hmt_{p_2}\gga+\ff{\eta}{2}C*\hmt_{p_1}\gga*C=
-\ff{\eta}{2}C*C*(\hmt_{p_2}-\hmt_{p_1+2p_2})\gga\,.
\ee
Here $p_1$ and $p_2$ differentiate first and second $C$'s
correspondingly as seen from left.
%Recall our convention $p_{\al}=-i\p^{C}_{\al}$.
Its solution within some $q$-homotopy \eqref{homs} reads
\be\label{B2q}
B_2:=B_2^{q}=\ff{\eta}{4i}C*C*\hmt_{q}(\hmt_{p_2}-\hmt_{p_1+2p_2})\gga\,,\quad
q=v_1p_1+v_2p_2\,.
\ee
Postponing the analysis of the freedom associated with $y$-shifts
in $q$ till section \ref{locfr} let us take $q$ \eq{hommn} (note
that the difference in the position of $\hmt_{q}$ in \eq{hommn}
and \eq{B2q} is equivalent to the shift $v_i\to v_i +1$ that, as
shown below, does not affect the final result).

Using \eq{hmtd} and \eq{unit} along with the fact that $ \hmt_a\hmt_b\hmt_c\gga =0$ $\forall a,b,c$
\eq{B2q} gives
\be\label{b2h}
B^{q}_2=\ff{\eta}{4i}C*C*(1-h_{q})\hmt_{p_1+2p_2}\hmt_{p_2}\gga\,.
\ee
Thanks to \eqref{h0}, for $q$ \eq{hommn},
\be
h_{q}\hmt_{p_1+2p_2}\hmt_{p_2}\gga =0
\ee
provided that $v_2-v_1=1$ which is just the PLT condition. Thus,
in this case the final result is independent of the remaining
freedom in $v_{1,2}$ giving
\be\label{b2}
B_2=\ff{\eta}{4i}C*C*\hmt_{p_1+2p_2}\hmt_{p_2}\gga\,.
\ee
As will be shown in the next section, these parameters however
contribute in presence of $y$-dependent shifts affecting local
terms. In accordance with PLT of \cite{Gelfond:2018vmi}  $B_2$
\eq{b2} must result in the local vertex.

From \eqref{HS3} we have at given order
\be
\dr_x C+[\go,C]_*+\dr_x B_2+[\go, B_2]_*+[W_1, C]_*=0\,.
\ee
Substituting here $B_2$  \eqref{b2} and $W_1$ from \eqref{W1om},
after simple algebra using \eqref{drag} and \eqref{gammaid1} one
eventually finds
\be\label{C2gen}
\dr_x C+[\go,C]_*=%\Upsilon(\go,C,C)+c.c.\,\qquad \Upsilon(\go,C,C)=
 \Upsilon_{\go
CC}+\Upsilon_{CC\go}+\Upsilon_{C\go C}+c.c.\,,
\ee
where
\begin{align}
\ls&   \Upsilon_{\go CC}=\ff{\eta}{4i} \go\!*\!C\!*\!C\!*\!X_{\go
CC} ,\quad
\Upsilon_{CC\go}=\ff{\eta}{4i}C\!*\!C\!*\!\go\!*\!X_{CC\go}
\,,\quad\Upsilon_{C\go C}=\ff{\eta}{4i}C\!*\!\go\!*\!C\!*\!X_{C\go C}\,,  \label{UPSWCC}\\
&X_{\go
CC}=h_{p_2}\hmt_{p_1+2p_2}\hmt_{p_1+2p_2+t}\gga\,,\label{XC1loc}\\
&X_{CC\go}=h_{p_2+2t}\hmt_{p_2+t}\hmt_{p_1+2p_2+2t}\gga\,,\label{XC2loc}\\
&X_{C\go
C}=(h_{p_1+2p_2+2t}-h_{p_2})\hmt_{p_2+t}\hmt_{p_1+2p_2+t}\gga\label{XC3loc}\,.
\end{align}
Local structures \eqref{XC1loc}-\eqref{XC3loc} are independent of
the leftover parameter \eqref{C2}. Note also that vertices
$\Upsilon(\go, C,C)$ are built from $h_a\hmt_b\hmt_c\gga$
structures related to the triangle function \eqref{tride} via
\eq{trf}.

It is straightforward to compute $\Upsilon(\go, C,C)=
\Upsilon_{\go CC}+\Upsilon_{CC\go}+\Upsilon_{C\go C}$ \eq{UPSWCC}
that enters \eqref{ver2} using \eqref{AdefHhhgam}. The final
result is
 \begin{align}\label{goCC}
\Upsilon_{\go CC}=&\ff{\eta}{2i}\int_{[0,1]^3}d^3\tau
\gd(1-\tau_1-\tau_2-\tau_3)
(\p_{1}^{\al}+\p_{2}^{\al})\p^{\go}_{\al} \\
& \go((1-\tau_3)y,\bar{y})\,\overline{*}\,
C(\tau_1y-i(1-\tau_2)\p^{\go},\bar{y};K)\,\overline{*}\,C(-(1-\tau_1)y+i\tau_2\p^{\go},\bar{y},K)k\,,\notag
\\\label{CCgo}
\Upsilon_{CC\go}=&\ff{\eta}{2i}\int_{[0,1]^3}d^3\tau
\gd(1-\tau_1-\tau_2-\tau_3)(\p_{1}^{\al}+\p_{2}^{\al})\p^{\go}_{\al} \\
&
C((1-\tau_3)y-i\tau_1\p^{\go},\bar{y};K)\,\overline{*}\,C(-\tau_3
y+i(1-\tau_1)\p^{\go},\bar{y},K)\,\overline{*}\,\go(-(1-\tau_2)y,\bar{y})k\,,\notag
\\\label{CgoC}
\Upsilon_{C\go C}=&\ff{\eta}{2i}\int_{[0,1]^3}d^3\tau
\gd(1-\tau_1-\tau_2-\tau_3)
(\p_{1}^{\al}+\p_{2}^{\al})\p^{\go}_{\al} \\\nn
&\Big\{C((1-\tau_3)y+
i(1-\tau_2)\p^{\go},\bar{y};K)\,\overline{*}\,
\go(\tau_1y,\bar{y})\,\overline{*}\, C(-\tau_3y-i\tau_2\p^{\go},\bar{y};K)\\
&+ C(\tau_2
y+i\tau_3\p^{\go},\bar{y};K)\,\overline{*}\,\go(-\tau_1y,\bar{y})\,\overline{*}\,
C(-(1-\tau_2)y-i(1-\tau_3)\p^{\go},\bar{y};K)\Big\}k\,,\notag
\end{align}
where $\p_{1,2}$ differentiates $C$'s. The result generalizes
local vertex $\Upsilon(\Omega, C, C)$ obtained in
\cite{Vasiliev:2016xui}  and remains local at cubic level.
Let us  stress that the remaining star product $C\bar{*}C\sim
e^{i\bar{p_{1}}^{\al}\bar{p_2}_{\al}}$ does not imply
non-locality. Indeed, since the vertex $O(CC)$ is
 local in its holomorphic part the contribution from the
anti-holomorphic one automatically terminates when restricted to
given three spins $s_1,s_2,s_3$.
Relaxing \eqref{homb} makes the corresponding vertex non-local. In
Appendix it is demonstrated that no parameters $v_1$ and $v_2$ in
\eqref{B2q} lead to a local vertex $\Upsilon(\go, C, C)$ other
than those specified by \eqref{C2}.

\section{$Y$-dependent shifts and local freedom}\label{locfr}
Perturbation theory  based on generalized resolution \eqref{homs}
is shown to properly reproduce simplest local vertices
\eqref{gogoC}-\eqref{goCgo} and \eqref{goCC}-\eqref{CgoC}. In this
analysis the shift parameters $q_i$ of $\hmt_{q_i}$ were
 linear combinations of derivatives $\p_i$ acting on the fields $C$
and $\go$. The freedom in relative coefficients turned out to be
constrained by locality implemented via PLT that keeps track of
derivative shifts in homotopies \eqref{hom}. Generally, however,
one can add  $y$-terms in $q_i$ as in, {\it e.g.,}
\eqref{dg}. The effect they produce is local at given order and,
naively, can be discarded. Indeed, thanks to the  star-product exchange
formulas \eqref{lpr} and \eqref{rpr} the resolution operators
\eqref{homs} can be arranged to act on $\gga$ only. As a result,
 the $y$-contribution  resides solely in the
pre-exponential as in (\ref{dg}).
It should be stressed however that the freedom in  local
field redefinition may affect the structure of
non-localities at higher orders. Therefore, we would like to control
 local $y$-shifts in \eqref{homs} aiming at future
higher-order application.

\subsection{Uniform $y$-shift freedom of $\Upsilon(\go,\go, C)$}

In our analysis we have shown that standard approach based on the
conventional homotopy used in \eqref{S1om} and \eqref{W1om} not
only reproduces canonical form of the central on-mass-shell
theorem but also leads to its ultra-local completion
\eqref{gogoC}-\eqref{goCgo}. Remarkably, there exists a
one-parameter extension of the conventional homotopy that
leaves vertex \eqref{gogoC}-\eqref{goCgo} unaffected.

To obtain it similarly to \eqref{S1om} and \eqref{W1conv} taking into account \eq{lpr} one can
solve for $S_1$ and $W_1$ in the following form
\begin{align}
 S_1&=-\ff{\eta}{2}\hmt_{\al(p+ y)}(C*\gga)+c.c.=-\ff{\eta}{2}C*\hmt_{p+\al y}\gga+c.c.\,,\label{S1al}\\ W_1&=\,\,\ff{1}{2i}\hmt_{\al(p+ y)}(\dr_x S_{1}+\go *S_1 + S_1*\go)+c.c.\,\nn\\&=-\ff{\eta}{4i}\left( C*\go*\hmt_{p+t+\al y}\hmt_{p+2t+\al
y}\gga- \go*C*\hmt_{p+t+\al y}\hmt_{p+\al
y}\gga\right)+c.c.\,.\label{W1al}
\end{align}
Homotopies, that contain one and the same coefficient in
$y$-shifts ($\al$ in \eqref{S1al} and \eqref{W1al}),
 will be referred to  as  {\it uniformly shifted}. That \eqref{S1al} and \eqref{W1al}
still reproduce \eqref{gogoC}-\eqref{goCgo} is a consequence of
identity \eqref{idy}, which guarantees that \eqref{X1}-\eqref{X3}
while receiving uniform $y$-shifts in homotopy parameters remain
unchanged. $\al=0$ corresponds to the conventional homotopy in
\eqref{S1al}, \eqref{W1al} reproducing  \eqref{S1om}
and \eqref{W1om}. Let us stress that should one took non-uniform
shifts by taking different parameters, say, $\al_1$ in $S_1$ and
$\al_2$ in $W_1$ the resulting vertex $\Upsilon(\go,\go,
C)$ would have been different from that in
\eqref{gogoC}-\eqref{goCgo} affecting the form of the free equation \eq{oms}
which is not allowed.
Correspondingly, expressions \eqref{S1al} and \eqref{W1al} provide
a one-parameter family of homotopies that give the same form of
central on-mass-shell theorem \eqref{oms} as well as of  its
completion $\Upsilon(\go,\go, C)$ \eqref{gogoC}-\eqref{goCgo}. Note that  $Y$-shifts,
uniform or not,
do not affect the class of homotopies \eq{homb} prescribed by PLT \cite{Gelfond:2018vmi}.

\subsection{Uniform $y$-shift freedom of $\Upsilon(\go,C, C)$}

Consider now the second-order contribution to the zero-form sector
of $B$ using the $\al$-modified uniform homotopies. Analogously to
the one-form sector we can directly generalize \eqref{b2} by the
uniform $y$-shift. Choosing appropriately shifted homotopies from
\eqref{S1al} and \eqref{B2} we find
\be\label{b2alfa+}
B_2=\ff{\eta}{4i}C*C*\hmt_{p_1+2p_2+\al y}\hmt_{p_2+\al
y}\gga\,.
\ee
While the obtained $B_2$ differs from  \eqref{b2}
 at $\al\neq0$ the physical sector
equation \eqref{C2gen} remains unchanged thanks to identity
\eqref{idy},  reproducing same vertices
\eqref{goCC}-\eqref{CgoC} as at $\al=0$.

Despite physical vertices $\Upsilon(\go,\go,
C)$ and $\Upsilon(\go, C,C)$ go through the uniform  homotopy
deformation this is no longer so for higher-order vertices
where  such a homotopy freedom may affect locality.
However  the non-uniform shifts  produce non-zero local effects in the lowest orders
which is interesting to inspect. Let us examine the zero-form sector.

\subsection{Non-uniform $y$-shift in $B_2$}\label{yshift}

In the previous section it was shown how shifted homotopy leads to the
expression for $B_2$ that differs from the one from
\cite{Vasiliev:2017cae}  by the $Z$-independent local term in
(\ref{der}).
 Adding a $y$ shift to the homotopy index it is possible
to obtain the additional local term.
Namely, starting from
the general expression for $B_2$
\begin{equation}
B_2=
\dfrac{i\eta}{4}C\ast C \ast\hmt_{\tilde{q}}
(\hmt_{p_1+2p_2}-\hmt_{p_2})
\gamma
\end{equation}
we specify index $\tilde{q}$
in the  form
\begin{equation}\label{q=}
\tilde{q}= q+\al y\q  q=v_1 p_1+v_2 p_2.
\end{equation}
Coefficients in front of $p_1$ and $p_2$ (recall the definition of
$p$  \eqref{p} ) are chosen to obey PLT condition \eqref{C2}
while $\al$ is  free.

  To proceed we  use  \eqref{ddg} whence it follows % rewritten in the form
 \bee\label{ddgy}
\hmt_{b+\al y}\hmt_a \gamma \!&=&\! 2\int_{[0,1]^3}   {d^3 \tau\,}
\gd(1-\tau_1-\tau_2-\tau_3)(z+b+ \al y)_\gga
 (  a-b-\al y)^\gga e^{i (\tau_1 z -\tau_2 a-\tau_3 b)_{\al}y^\ga}k\qquad\\ \nn\!&=&\!
 \hmt_{b }\hmt_a \gamma+2\int_{[0,1]^3}   {d^3 \tau\,}
\gd(1-\tau_1-\tau_2-\tau_3) (z+a)_\gga
 (  -\al y)^\gga  e^{i (\tau_1 z -\tau_2 a-\tau_3 b)_{\al}y^\ga}k
\\ \nn\!&=&\!
  \hmt_{b }\hmt_a \gamma +2i\al\int_{[0,1]^3}   {d^3 \tau\,}
\gd(1-\tau_1-\tau_2-\tau_3) \left(\frac{\p}{\p \gt_1}- \frac{\p}{\p \gt_2}\right)   e^{i (\tau_1 z -\tau_2 a-\tau_3 b)_{\al}y^\ga}k.\eee
By partial integration  this yields
\bee\label{ddgy=}
\hmt_{b+\al y}\hmt_a \gamma=  \hmt_{b }\hmt_a \gamma -2i\al\int_{[0,1]^3}   {d^3 \tau\,}
\gd(1-\tau_1-\tau_2-\tau_3) \left(\gd(   \gt_1)- \gd(\gt_2)\right)   e^{i (\tau_1 z -\tau_2 a-\tau_3 b)_{\al}y^\ga}k.\eee
Therefore by virtue of \eq{ddgy=}
from   \eqref{B2q} it follows for $\tilde{q}$ \eq{q=}
  \bee  \label{p12p2}
 B_2^{\tilde{q}}=
\dfrac{i\eta}{4}C\ast C \ast\hmt_{q+\al y}
(\hmt_{p_1+2p_2}-\hmt_{p_2})
\gamma=\\ \nn= B_2^{q}
+
  \dfrac{\eta  \al}{2}C\ast C \ast \int_{[0,1]^2}   {d^2 \tau\,}
\gd(1 -\tau_2-\tau_3) \left(
e^{i (  -\tau_2 ({p_1+2p_2})-\tau_3 {q})_{\al}y^\ga}
-e^{i (  -\tau_2 { p_2}-\tau_3 {q})_{\al}y^\ga}
\right)k
 \eee

  As anticipated, the $y$-shift in $\hmt_{q+\alpha y}$
   does not affect the exponential and hence locality.
  As
mentioned in section \ref{vert}  the   $\ga$--independent part $B^{q}_2$
 gives $B_2$ of \cite{Vasiliev:2017cae}
with the local extra term.

Choosing appropriately $q$ and $\al$ one can further simplify
the extra term on the r.h.s. of \eqref{der}. The  two most interesting options are
\begin{equation}
\label{B2L} q=p_1+2p_2,\;\; \al=-1\,,
\end{equation}
\begin{equation}
\label{B2R} q= p_2,\;\; \al=1\,.
\end{equation}

In the case  \eqref{B2L} the   $\al$-dependent terms from \eqref{p12p2}
  can be rewritten in the following way
\bee\label{shiftL}
  -\dfrac{\eta   }{2}C\ast C \ast \int_{[0,1]^2}   {d^2 \tau\,}
\gd(1 -\tau_2-\tau_3) \left(
e^{  -  y^\al (\p_1+2\p_2)_\al}
-e^{-y^\al \p_{2\al}-\tau_3 y^\al (\p_1+\p_2)_\al}
\right)k.
 \eee
Straightforward computation of this expression gives
\begin{equation}
-\frac{\eta}{2}C\left(0,\bar{y};K\right) \bar{\ast}\, C\left(-y,\bar{y};K\right)k+\frac{\eta}{2}\int_0^1 dt
\,C\left(ty,\bar{y}; K\right)\,\overline{*}\,C\left(\left(t-1\right)y,\bar{y};
K\right)k,
\end{equation}
which yields by virtue of \eq{der}
\begin{equation}\label{FL}
\frac{i\eta}{4}C\ast C \ast
\hmt_{p_1+2p_2-y}\left(\hmt_{p_1+2p_2}-\hmt_{p_2}\right)\gamma=B^{loc}_{2\eta}-\frac{\eta}{2}C\left(0,\bar{y};K\right) \bar{\ast}\, C\left(-y,\bar{y};K\right)k.
\end{equation}
Analogously,
in the case  \eqref{B2R} the   expression for $B_2$ can be written in the form
\begin{equation}\label{FR}
\frac{i\eta}{4}C\ast C \ast
\hmt_{p_2+y}\left(\hmt_{p_1+2p_2}-\hmt_{p_2}\right)\gamma=B^{loc}_{2\eta}-\frac{\eta}{2}C\left(y,\bar{y};K\right)\bar{\ast}\, C\left(0,\bar{y};K\right)k.
\end{equation}

A natural choice for $B_2$  respecting the symmetry  reversing the order of
product factors is
\begin{multline}
B_2=\frac{i\eta}{8}C\ast C \ast\left( \hmt_{p_1+2p_2-y}+\hmt_{p_2+y}\right)\left(\hmt_{p_1+2p_2}-\hmt_{p_2}\right)\gamma=\\
=B^{loc}_{2\eta}-\frac{\eta}{4}C\left(0,\bar{y};K\right) \bar{\ast} k\, C\left(y,\bar{y};K\right)-\frac{\eta}{4}C\left(y,\bar{y};K\right)\bar{\ast}k\, C\left(0,\bar{y};K\right)\,.
\end{multline}

{
 Using shifted homotopy Ansatz  we were not able to reproduce
$B^{loc}_{2\eta}$  with no additional  local terms. However the
terms  on the $\rhs{}'s$ of \eqref{FL} and \eqref{FR} contributing
to dynamical equations feature quasi ultra-local form, \ie such
that the argument of one of the factors of $C$ is zero and there
is at most a finite number of  contractions between them. }

Also note that in presence of the
 $y$-shift the final result does depend on $v_1+v_2$ which freedom only affects
 local terms.
 \section{Conclusions}\label{con}

The important problem addressed in \cite{Gelfond:2018vmi} and in this paper is the elaboration of the technical
tools allowing to find a minimally nonlocal formulation of the HS theory
and its further generalizations like those proposed recently in
\cite{Vasiliev:2018zer}. The main new tool consists of the application of the
shifted homotopy techniques allowing to reduce the degree of nonlocality
of HS vertices in accordance with the Pfaffian Locality Theorem (PLT) of \cite{Gelfond:2018vmi}
(see also Section \ref{Gelfond}). In this paper we have applied this approach to the
analysis of some vertices in HS theory demonstrating that the shifted homotopy from
the proper class identified in \cite{Gelfond:2018vmi} leads directly to the known local vertices
of \cite{Vasiliev:2016xui,Vasiliev:2018zer} with no need of nonlocal field redefinitions
used in those references.

In this paper we develop perturbative approach to HS equations
\eqref{HS1}-\eqref{HS5} based on shifted resolutions \eqref{homs}
$\hmt_q$ generalizing the conventional resolution $\hmt_0$
introduced in \cite{more} that is known to render non-localities
in cubic vertices \cite{GY1} (see also \cite{Boulanger:2015ova}).
Parameter $q$ in its argument is any $z$-independent spinor
variable (operator). The proposed homotopies drastically simplify
the analysis of HS equations thanks to remarkable star-product
exchange formulas \eqref{lpr}, \eqref{rpr} found in this paper,
which link together such seemingly unrelated operations as star
product and homotopy integration. They reduce the analysis of HS
equations to the analysis of structures built from combinations of
resolutions $\hmt_{a_i}$ acting on $\gamma$ and their further star
products,
 where $\gga$ is the  central element
\eqref{klein} from HS equations. Though at higher orders  these
structures get more and more involved like
$\hmt_a\hmt_b\gga*\hmt_c\hmt_d\gga$, $\hmt_a(\hmt_b\gga*\hmt_c\gga)$, {\it etc},
 they remain the only
ones\footnote{Strictly speaking, this is true up to
 terms resulting from the application of $\dr_x$ to dynamical fields $\go$ and $C$
 in the lower orders, which  also contain  the cohomology projectors
 $h_Q$ \eq{hf} like in \eq{C2gen}, \eq{UPSWCC}, \eq{XC1loc}-\eq{XC3loc}, that
 demand a proper extension of the shifted homotopy technique to be presented
 elsewhere.} to be analyzed leading to enormous simplification of the
perturbative analysis. The freedom in homotopy operator parameters
reflects the freedom in the gauge choice and field redefinition of
dynamical fields and therefore is tightly related to the locality
problem. One of the key results of this paper is that PLT that
constrains the homotopy freedom leaves one with the family of
parameters resulting in $\Upsilon(\go,\go, C)$ and
$\Upsilon(\go,C,C)$ being ultra-local and local correspondingly
for an arbitrary HS connection $\go$. Our results can be
summarized as follows.

\begin{itemize}
\item One of the major and simple results of the paper are
star-product exchange formulas \eqref{lpr}, \eqref{rpr} that relate
shifted homotopies and the star product. The obtained expressions
are largely rely on the form of the star product
\eqref{star}. By making it possible to interchange star product of
HS fields with homotopy operation it shifts gears of the whole
perturbation theory. The new setting reduces it eventually to the
analysis of homotopies and their star products of the central elements
$\gamma$ and $\bar \gamma$
 only. In particular, it allows one analyzing HS vertices
in generic backgrounds.

\item Using the notion of shifted homotopies \cite{Gelfond:2018vmi} we have
studied simplest HS vertices $\Upsilon(\go,\go,C)$ and
$\Upsilon(\go,C,C)$. In doing so we have modified the standard
perturbative approach that is based on the expansion around $AdS$
background to the HS curvature $C$-expansion allowing one directly
reproducing vertices in \eqref{ver1}, \eqref{ver2}. From this
point of view the found vertices represent cubic completion of
Central on-shell theorem \eqref{oms} and quadratic vertex
$\Upsilon(\Omega_{AdS},C,C)$. By analyzing homotopy parameter
space we check explicitly that a subspace that renders dynamical
equations local is that  governed by the PLT.

\item We have also briefly addressed the problem of local field
redefinitions. Within the proposed generalized set of homotopies
$\hmt_q$ there are two different kind of parameters encoded in
$q$. One is the shift in  derivatives   over the arguments of fields $C$ and
  $\go$ and the other is just $y$'s. The former
crucially affect local structure of the resulting dynamical
equations. We show that in agreement with PLT there is a  family
of shift parameters that make
equations local. On the other hand,
shifts $q\sim y$ at the lower level of perturbation theory are
responsible for local effects remaining unconstrained by
locality. It is important however to control
local shifts as well as they can affect locality at higher orders.

{ In doing so we found a one-parameter family of homotopies
consistent with the admissible class of \cite{Gelfond:2018vmi}
which while affecting HS master fields in their perturbative
expansion provides the same dynamical equations. Particularly,
this family respects Central on-shell theorem. That freedom is not
expected to leave higher order vertices unaffected and therefore
provides a one parametric control over higher-order structures.}
\end{itemize}

Let us note that the analysis carried out in this paper is mostly
focused on the holomorphic sector of the HS equations (at this
level the anti-holomorphic sector is reproduced
analogously). As such it can be directly used for analysis
of $3d$ HS equations.

There are plenty of problems left outside the scope of the paper.
Particularly, we have considered the simplest HS vertices governed
by deformation of HS algebra. These turned out to be (ultra)
local. Among those bilinear in  $C$  there is one,
$\Upsilon(\go,\go, C,C)$, which
we left aside. This vertex being sufficiently simple is crucial
for HS locality as it contains structures typical for higher
orders. Namely, unlike vertices considered in this paper it has
star product of homotopy structures, e.g.,
$\hmt\hmt\gga*\hmt\hmt\gga$. It will be interesting to see whether
it is local or not. What is known at this stage is that its reduction
to $AdS$, $\Upsilon(\Omega, \Omega, C, C)$ should be local being a
part of HS cubic vertex found in \cite{Gelfond:2017wrh}. This  does not
necessarily imply that $\Upsilon(\go,\go, C,C)$  is local for general $\go$.

Another key test for (non-)locality is the $\Upsilon(\go, C,C,C)$
vertex. At the level of vertices considered in this paper the
notions of space-time locality and spin locality from our analysis
are equivalent. At higher orders however the two may essentially
differ. Vertex $\Upsilon(\go, C,C,C)$ in this context is of
particular interest given the result of \cite{Bekaert:2015tva},
where its $AdS$ reduction was holographically analyzed for scalar
quartic self-interaction. Interestingly, its space-time
representation presumably is non-local but have non-locality of a
specific form \cite{Ponomarev:2017qab}.

Last but not the least, it will be interesting to understand the
role of PLT at higher orders. At the level of
vertices considered in our paper PLT prescribes unique
(up to local redefinitions) form of HS vertices. Nonetheless this
theorem looks so far as a crude estimate on exponential behavior
in perturbation theory. What it says is that the class of
homotopies in the one-form sector of HS equations is respected
provided the zero-form homotopy sector is appropriately shifted
\eqref{hom}, \eqref{homb}. The key observation
was that the inner Klein operator present in HS equations breaks down the
class of functions generated by star products along with homotopy
action unless that breaking is taken into account by properly
shifted homotopies in zero-form sector. Strikingly, this is what
is needed to come up with local lower order vertices. PLT
however says nothing about concrete values of homotopy
coefficients in \eqref{hom} other than single constraint
\eqref{homb}. It will be interesting to see if PLT admits a
generalization strong enough to fully control over admissible structure
of higher order vertices.

The important current problem in
HS theory is to understand to which extent it is local.
If non-local then one should elaborate a proper
substitute for the notion of locality (minimal non-locality),
{\it i.e.} the admissible functional
class that respects physical predictions be these HS correlation
functions, black hole solutions, charges, {\it etc}.
 The study of  systematic means for
addressing these kind of questions is initiated in \cite{Gelfond:2018vmi}
and  in this paper.

\section*{Acknowledgments}
VD is grateful to Mitya Ponomarev for valuable correspondence. OG
and MV are grateful to Ivan Degtev for the independent check that
vertex \eq{go1} properly reproduces Central-on-shell Theorem in
$AdS_4$. We would also like to thank Oleg Shaynkman for careful
reading the paper and useful remarks. This work was supported by
the Russian Science Foundation grant 18-12-00507.

\section*{Appendix}

 Here we analyse in some more detail the structure of
$\Upsilon(\go,C,C)$ for \eqref{B2q} with unconstrained
coefficients $v_1$ and $v_2$ to show that such vertex is
non-local unless \eqref{C2} imposed. Straightforward calculation yields
\be\label{C2gen1}
\dr_x C+[\go,C]_*=%\Upsilon(\go,C,C)+c.c.\,\qquad \Upsilon(\go,C,C)=
 \Upsilon_{\go
CC}+\Upsilon_{CC\go}+\Upsilon_{C\go C}+c.c.\,,
\ee
where
\begin{align}
\ls&   \Upsilon_{\go CC}=\ff{\eta}{4i} \go\!*\!C\!*\!C\!*\!X_{\go
CC} ,\quad
\Upsilon_{CC\go}=\ff{\eta}{4i}C\!*\!C\!*\!\go\!*\!X_{CC\go}
\,,\quad\Upsilon_{C\go C}=\ff{\eta}{4i}C\!*\!\go\!*\!C\!*\!X_{C\go C}\,, \\
&X_{\go
CC}=h_{v_1p_1+v_2p_2}\hmt_{p_1+2p_2}\hmt_{p_1+2p_2+t}\gga\!-\!(h_{v_1p_1+v_2p_2}
\!-\!h_{v_1p_1+v_1t+v_2p_2})\hmt_{p_2}\hmt_{p_1+2p_2+t}\gga\,,\label{XC1}\\
&X_{CC\go}=h_{v_1p_1+v_2p_2+2t}\hmt_{p_2+2t}\hmt_{p_1+2p_2+2t}\gga\!-\!
(h_{v_1p_1+v_2p_2+v_2t}\!-\!h_{p_2+2t})\hmt_{p_2+t}\hmt_{p_1+2p_2+2t}\gga\,,\label{XC2}\\
&X_{C\go
C}=(h_{p_2+t}\!-\!h_{v_1p_1+v_1t+v_2p_2})\!\hmt_{p_2}\hmt_{p_1+2p_2+t}\gga+(h_{p_1+2p_2+t}
\!-\!h_{v_1p_1+v_2p_2+v_2t})\!\hmt_{p_1+2p_2+2t}\hmt_{p_2+t}\gga\label{XC3}\,.
\end{align}
To see if the \rhs  of \eqref{C2gen1} is local or not, consider
typical expression that appears there
\be\label{homac}
C*C*h_{a_1p_1+a_2p_2}\hmt_{b_1p_1+b_2p_2}\hmt_{c_1p_1+c_2p_2}\gga\,,
\ee
where $a_{1,2}$, $b_{1,2}$ and $c_{1,2}$ are some numbers. We
discard the $\go$--factor and corresponding homotopy
differentiation parameter $t=-i\p^{\go}$ as it does not affect
locality being at most polynomial in $y$'s for a given spin. Now,
$C*C$ itself contains non-local contribution
\be
C*C\sim e^{ip_{1}^{\al}p_{2\al}}
\ee
which should be cancelled out by homotopy action \eqref{homac}.
Performing star product in \eqref{homac} using \eqref{ddg} and
keeping track of the non-local part we find it to be
\be
 (C\,\overline{*}\,C)
\,e^{i(1-(a_2-a_1)\tau_1-(b_2-b_1)\tau_2-(c_2-c_1)\tau_3)p_{1}^{\al}p_{2\al}}\,.
\ee
Due to integration over simplex \eqref{ddg} which cuts off
$\tau_1+\tau_2+\tau_3=1$, the exponential containing
$p_{1}^{\al}p_{2\al}$ cancels out iff
\be\label{locgen}
a_2-a_1=b_2-b_1=c_2-c_1=1\,.
\ee

It is easy to see now that all structures in
\eqref{XC1}-\eqref{XC3} meet locality requirement \eqref{locgen}
provided that \eq{C2} is true, \ie the homotopy for $B$  is taken
within the class of functions  \eqref{homb} specified for
zero-forms. This result is anticipated being in accordance with
PLT of \cite{Gelfond:2018vmi} stating that the Pfaffian matrix of
derivatives must be degenerate for the proper class of homotopies.
Being a $2\times 2$ matrix for the terms bilinear in  $C$ this
implies that it is zero.

\end{document}